\documentclass[amsfonts,amssymb,amsmath,prb,aps,reprint,notitlepage,superscriptaddress,longbibliography]{revtex4-1}
\usepackage[T1]{fontenc}
\usepackage{times}
\usepackage{xcolor}
\usepackage[colorlinks,allcolors=blue]{hyperref}
\usepackage{graphicx}
\usepackage{bm}
\usepackage{mathrsfs}
\usepackage{dcolumn}
\usepackage{float}
\usepackage[normalem]{ulem}
\usepackage{dsfont}
\usepackage{amsmath}
\usepackage{amsfonts}
\usepackage{amssymb}
\usepackage{braket}
\usepackage{stmaryrd}

\renewcommand{\vec}[1]{{\ensuremath{\bm{\mathrm{#1}}}}}

\newcommand{\muB}{{\ensuremath{\mu_{\mathrm{B}}}}}

\newcommand\ddfrac[2]{{\ensuremath{\frac{\displaystyle #1}{\displaystyle #2}}}}

\begin{document}

\title{Numerical Simulations of a Spin Dynamics Model Based on a Path Integral Approach}

 \author{Thomas Nussle}
\email{t.s.nussle@leeds.ac.uk}
\affiliation{School of Physics and Astronomy, University of Leeds, Leeds, LS2 9JT, United Kingdom}
 \author{Stam Nicolis}
\email{stam.nicolis@lmpt.univ-tours.fr}
\affiliation{Institut Denis Poisson, Université de Tours, Université d'Orléans, CNRS (UMR7013), Parc de Grandmont, F-37200, Tours, France}
 \author{Joseph Barker}
\email{j.barker@leeds.ac.uk}
\affiliation{School of Physics and Astronomy, University of Leeds, Leeds, LS2 9JT, United Kingdom}

\begin{abstract}
Inspired by path integral molecular dynamics, we build a spin model, in terms of spin coherent states, from which we can compute the quantum expectation values of a spin in a constant magnetic field, at finite temperature.
This formulation facilitates the description of a discrete quantum spin system in terms of a continuous classical model and recasts the quantum spin effects within the framework of path integrals in a double $1/s$ and $\hbar s$ expansion, where $s$ is the magnitude of the spin. In particular, it allows for a much more direct path to the low- and high-temperature limits of the quantum system
and to the definition of effective classical Hamiltonians that describe both thermal and quantum fluctuations.
In this formalism, the quantum properties of the spins emerge as an effective anisotropy. We use atomistic spin dynamics to sample the path integral, calculate thermodynamic observables and show that our effective classical models can reproduce the thermal expectation values of the quantum system within 
temperature ranges relevant for studying magnetic ordering.
\end{abstract}

\maketitle

\section*{Introduction\label{introduction}}
Spin models of magnetic materials are usually either quantum or classical in terms of the elementary building blocks on which they are based. In quantum spin models, the spin states belong to the quantum space of states that includes all linear superpositions of the eigenstates of $\hat{S}_z$ and $\hat{\bm{S}}^2$, and the spin variables are quantum operators. By contrast, in classical spin models, `spin' is used colloquially and actually refers to the classical magnetic moment, $\mu\vec{S}$, where $\vec{S}$ is usually of fixed length with dynamics confined to the surface of the Bloch sphere and $\mu$ is the spin magnetic moment in Bohr magnetons.

Quantum models allow an accurate description of both thermodynamics and dynamics, 
which intrinsically include purely quantum effects such as entanglement and quantum fluctuations. However, the size of systems that can be studied is often limited to tens or hundreds of spins due to the large computational cost, as solving quantum problems exactly amounts to diagonalization of larger and larger matrices, and even approximation schemes thereof suffer from scaling issues. Numerical methods, such as quantum Monte Carlo (QMC), allow calculations of very large quantum spin systems (hundreds of thousands of spins) with very high accuracy. However, there is no access to dynamical quantities, as QMC is intrinsically a description of thermodynamics, where time is absent. Other quantum methods which do provide access to real-time dynamics cannot provide results for such large systems. Additionally, fundamental issues also arise, such as the `sign problem' in the case of antiferromagnets, since the Hubbard-Stratonovich transformation leads to an effective Hamiltonian that is not hermitian although the evolution operator is unitary~\cite{Ceperley_Science_231_555_1986}. 

Classical spin models are frequently used to study the dynamics and thermodynamics of magnetic materials, helping to interpret experiments at ``high'' temperatures, where quantum effects-such as entanglement-can be neglected. The computational cost is relatively low, and the formalism is easy to parallelize, leading to routine simulations of the dynamics of hundreds of thousands or even millions of spins. While these classical models give a good qualitative description of the magnetic dynamics, issues arise at lower temperatures, where the assumption of classical Boltzmann statistics is no longer appropriate. The magnon Debye temperature tends to be very high and of the same order
as the magnetic ordering temperature, so the `low-temperature' regime may cover most of the temperature range of magnetic ordering~\cite{Barker_PhysRevB_100_140401_2019, Barker_ElectronStruct_2_044002_2020}. Recent efforts have been made to introduce ad hoc corrections to classical spin models to produce results that more closely resemble quantum models and to better agree with experimental measurements \cite{Woo_PhysRevB_91_104306_2015, Bergqvist_PhysRevMater_2_013802_2018,Evans_PhysRevB_91_144425_2015, Barker_PhysRevB_100_140401_2019, Anders_NewJPhys_24_033020_2022, Walsh_npjComputMater_8_186_2022}. However, these approaches are incapable of including quantum effects, such as tunneling between macroscopic states or zero-point fluctuations. These quantum effects are becoming relevant at ever
larger length scales and higher temperatures, for example, with the measurement of the motion of domain walls induced by quantum domain fluctuations in Cr up to 40K~\cite{Shpyrko_Nature_447_68_2007}. Thus, what is still lacking is a dynamical quantum model whose accuracy can bridge the gap between a fully quantum simulation of a few atoms and an effective classical model and that enables simulations scalable to the size of spintronic device components of millions of spins.

Here, we describe a way of constructing
a bridge between quantum and classical spin models by employing a path integral formalism for spin dynamics. This is inspired by path integral molecular dynamics~\cite{Parrinello_JChemPhys_80_860_1984} where the efficiency of classical molecular dynamics is used to calculate quantum properties, by establishing the appropriate evolution equations to move in the phase space of the quantum system and thus sample configurations therein~\cite{Habershon_AnnuRevPhysChem_64_387_2013}. However, how to take into account spin degrees of freedom and sample the corresponding phase space is by no means obvious. 

First attempts to do so~\cite{Runeson_JChemPhys_152_084110_2020}, in particular for molecular magnets~\cite{Coronado_NatRevMater_5_87_2019} express the spin degrees of freedom in terms of equivalent, though fictitious, position and momentum variables and using the known molecular dynamics formalism in this guise. Hence, these involve mapping the spin Hamiltonian to a particle Hamiltonian. This makes the interpretation of the results in terms of classical magnetic moments, the actual experimental observable, much less straightforward, and this mapping is difficult to build for more complex spin interactions. However, the real problem which we must overcome is that the space of positions and momenta is flat; while the space spanned by the spin degrees of freedom is curved.

It is this problem that is solved by using the basis of spin coherent states~\cite{Runeson_JChemPhys_152_084110_2020}. While spin coherent states have been used in some quantum methods~\cite{Bossion_JChemPhys_157_084105_2022}, these methods incur a non-trivial cost, for large systems, as well as not being well-suited for extracting the information on the individual (classical) spin components. 
We note, however, that  spin coherent states have also recently been used in methods to derive/rederive equations of motion for magnetization dynamics~\cite{Zhang_PhysRevB_104_104409_2021}.
Introducing spin coherent states comes at a price: these states are no longer eigenstates of the quantum Hamiltonian. Nonetheless, as we are interested in studying the crossover from quantum to classical behaviour, it is precisely these spin coherent states that are best suited for the task.

In this Article we therefore consider the simplest nontrivial spin system: a single spin in an external magnetic field, described by the Zeeman Hamiltonian. We develop a formalism which uses the spin coherent states and the operators that act on them to to compute themal expectation values of the quantum system in exactly solvable cases, and compare the results obtained to numerical calculations performed with classical atomistic spin dynamics methods, in presence of a field, 
which takes into account the quantum properties  of the spins, when in contact with a thermal bath. We demonstrate that this formalism can indeed account for the quantum properties of the spin, across a broad range of temperatures, with deviations appearing only at ``very low'' temperatures, as expected by intuition. We emphasise here that, we are not seeking an exact classical equivalent of the quantum system, rather, we are building an effective classical model whose thermal expectation values reproduce those of its quantum counterpart through a dynamical stochastic path sampling method. Moreover, the scope of this paper is not a fundamental study of path integrals for spin systems nor is it placed in the context of geometric quantisation schemes, although the literature from these fields has proven particularly useful for building our model and will be important in future works~\cite{kochetovQuasiclassicalPathIntegral1998,cabraPathIntegralRepresentation1997, klauderPathIntegralsStationaryphase1979}.

The plan of the paper is as follows: In Section \ref{section_1} we start from a description of a quantum spin system in terms of the discrete spin states $\ket{s,m}$ which are eigenvectors of $\hat{\bm{S}}_z$ and $\hat{\bm{S}}^2$ and switch to the continuous spin coherent state basis to show that from the quantum model, we can recover a continuous description which can be rewritten in terms of the classical spin vectors $\vec{S}$. We do this in a systematic double expansion in $1/s$ and $\hbar s$; To justify this we explicitly recover the classical limit from this formalism as a sanity check of our approach. In Section \ref{section_2} we consider special cases where results can be computed directly from the partition function. We compute expectation values using the spin coherent states for the classical limit (an exact result), for several orders of corrections to this classical limit (under our systematic approximation scheme) and for the exact quantum solution using the discrete $\ket{s,m}$ basis. These results serve as reference and are compared with the results obtained from the new method developed in the next section. In Section \ref{section_3} we begin by deriving an effective classical Hamiltonian from the quantum partition function in both low (Section \ref{lowTExp}) and high (Section \ref{high-t-model}) temperature limits. In both cases, the resulting effective classical magnetic system is sampled by computing stochastic paths on the Bloch sphere using finite temperature atomistic spin dynamics simulations. In fact, for the system at hand the path integral is an integral over the manifold of all possible superpositions, i.e. over a complex projective space. Finally, we compare results from classical atomistic spin dynamics simulations to results from our new enhanced atomistic model, whilst using the results obtained directly from the partition function (cf. Section \ref{section_2}) as reference. We show that indeed, we are able to recover the correct quantum thermal expectation values from this effective classical model for most of the temperature range where there is a significant difference between the classical limit and the quantum solution. In section \ref{conclusion}, we summarize our findings and discuss key issues to address in further work.

\section{From the spin states to the spin coherent states }\label{section_1}

\subsection{Partition function in the discrete spin states basis}

In molecular dynamics, the dynamical variables of the quantum system take values in a flat space. This makes the application of path integrals using classical positions and momenta relatively straightforward. For spin systems, the dynamical variables, the components of spin, take values in a curved space and can only take discrete values due to the discrete spectrum of the spin Hamiltonian
\begin{equation}
    \{\ket{s,m}\}\,,\quad m\in \llbracket -s,s \rrbracket,
\end{equation}
where $s$ is the principal quantum number and $m$ labels all different possible states with this given spin $s$. For example, with $s=2$ there are $2s+1=5$ eigenstates:
\begin{equation}
	\left\{\ket{2, -2},\ket{2, -1},\ket{2, 0},\ket{2, 1},\ket{2, 2}\right\}.
\end{equation}
However, all possible states of a quantum system of spin $s=2$ are linear combinations of these five states, i.e. they are described as 
\begin{equation}
\label{spin2moment}
|\psi\rangle=c_{-2}\ket{2, -2}+c_{-1}\ket{2, -1}+c_0\ket{2, 0}+c_1\ket{2, 1}+ c_2 \ket{2, 2}
\end{equation}
The normalization of these states implies that the coefficients satisfy the constraint
\begin{equation}
\label{norm_s=2}
|c_{-2}|^2+|c_{-1}|^2+|c_0|^2+|c_1|^2+|c_2|^2=1,
\end{equation}
which defines a point on the unit sphere in ten dimensions, but the property that five phases can be modded out reduces this to a five-dimensional manifold. The real challenge is to sample this space efficiently. 

The partition function of this quantum spin system is the volume of this five-dimensional manifold, which is finite:
\begin{equation}
\label{Zspin2}
\begin{array}{l}
\displaystyle
{\cal Z}=\int\,d^2c_{-2}d^2c_{-1}d^2c_0d^2c_1d^2c_2\,\\
\displaystyle
\delta(|c_{-2}|^2+|c_{-1}|^2+|c_0|^2+|c_1|^2+|c_2|^2-1).
\end{array}
\end{equation}
Upon coupling the magnetic moment to a thermal bath, the partition function takes the form
\begin{equation}
\label{Zspin2beta}
\begin{array}{l}
\displaystyle
{\cal Z} = \int d\psi\langle\psi|e^{-\beta H}|\psi\rangle=\\
\displaystyle
\int\,d^2c_{-2}d^2c_{-1}d^2c_0d^2c_1d^2c_2\,\\
\displaystyle
\delta(|c_{-2}|^2+|c_{-1}|^2+|c_0|^2+|c_1|^2+|c_2|^2-1)\,e^{-\beta H(c)},
\end{array}
\end{equation}
with $\beta=1/(k_B T)$, where $k_B=1.381\times10^{-23}$~J/K is the Boltzman constant and $T$ is the temperature in Kelvin. From Eq.~\eqref{Zspin2beta} it is not obvious how the dynamical behavior of the quantum system, defined over the full manifold, goes over to that of a classical system, localized on the five states $\left\{\ket{2, -2},\ket{2, -1},\ket{2, 0},\ket{2, 1},\ket{2, 2}\right\},$ in the ``classical limit'' and how this can be defined.

This requires a careful discussion of what we mean by a `quantum' system and its classical limit. On the one hand, we have the discrete basis of the eigenstates of the Hamiltonian, but on the other hand, we have the quantum superposition of states which leads to a continuous manifold of possible quantum states. Here, we emphasize that we are dealing with classical measurements of quantum systems, which means that the outcome of any single measurement can only be an eigenstate of our Hamiltonian-which is labeled by an integer for spin systems. The prototype of this situation is the experiment by Stern and Gerlach \cite{gerlachExperimentelleNachweisRichtungsquantelung1922}, where, even though the possible quantum states of the electron can belong to a superposition,
\begin{equation}
	\ket{\psi}=a\ket{\uparrow}+b\ket{\downarrow},
\end{equation}
such that $a^2+b^2=1$, the outcome of the measurement of the experiment is either $\ket{\uparrow}$ or $\ket{\downarrow}$. This is in contrast to a classical measurement of the projection along the $z$-axis of a classical magnetic moment for which a single measurement could take any value between $+\mu_s$ and $-\mu_s$ where $\mu_s$ is the total magnetic moment. Thus, if our Hamiltonian is a function of $\hat{S}_z$ only, then the partition function corresponding to the classical measurement of said quantum system is given as a sum over the eigenstates of this Hamiltonian, rather than an integral over the quantum manifold of states,
\begin{align}
\begin{split}
	{\cal Z} \equiv \mathrm{Tr}(e^{-\beta \hat{\cal H}})&=\sum_{m=-s}^{s}\bra{s,m}e^{-\beta \hat{\cal H}[\hat{S}_z]}\ket{s,m}\\
 &=\sum_{m=-s}^{s}e^{-\beta\hat{\cal H}[m]}
	\label{spinStatesPartitionFunction}
\end{split}
\end{align}
This expression can be evaluated, especially for the case of a single spin; however defining, let alone studying its classical limit is by no means obvious. It is to this end that it's useful to introduce the spin coherent states.

\subsection{Partition function in the continuous spin coherent state basis}

One way to sample the partition function over the quantum space of states,
that is particularly useful in studying the crossover to the classical limit,
is to recast the system in terms of the so-called spin coherent states~\cite{Radcliffe_JPhysAGenPhys_4_313_1971}. Indeed, not only do the spin coherent states form a continuous basis for the spin system, enabling a mapping onto the continuous description in terms of a unit vector living on a sphere, but it has also been shown that their behavior is close to the classical limit~\cite{Lee_Loh_AmJPhys_83_30_2015}.
Thus, they enable us to, on one hand efficiently sample the manifold of quantum states and other other hand to consistenly define the classical limit. The spin coherent states have previously been used to study fundamental aspects such as emerging supersymmetry in spin systems \cite{Stone_NuclPhysB_314_557_1989}, semiclassical transition probabilities \cite{Stone_JMathPhys_41_8025_2000}, and energy gap computations within mean-field quantum perturbation theory \cite{Koh_PhysRevB_97_094417_2018}.

We now proceed by introducing the spin coherent states and showing that the matrix elements of $\hat{S}_z$ can be written as a sum of the classical limit plus corrections. These corrections are essential for including quantum fluctuations into our effective model. We show results for both the purely classical limit of the spin coherent states and how the systematic inclusion of these corrections brings the expectation values closer to their quantum counterparts.

To use the spin coherent states, we work as follows: for a given quantum spin number $s,$ we set 
\begin{equation}
    \ket{p} \equiv \ket{s,s-p},
\end{equation}
where $p\in\{0,1,\ldots,2s-1,2s  \}$
using the labeling introduced above
and we define the spin coherent states $\ket{z},$ labeled by a complex number $z,$ by the action of the lowering operator~\footnote{Of course one can also define these in terms of the raising operator or any linear combination of these \cite{Nemoto_JPhysAMathGen_33_3493_2000}}, $\hat{S}_- =\hat{S}_x - i\hat{S}_y$, as
\begin{equation}
    \ket{z} \equiv(1+|z|^2)^{-s}\operatorname{e}^{z\hat{S}_-/\hbar} \ket{0}\label{coherentState}
\end{equation}
where the $1/\hbar$ factor is a bookkeeping device needed to keep the exponential dimensionless. 
Its role in setting the scale of the quantum fluctuations will emerge in what follows.
The action of $\hat{S}_+$, $\hat{S}_-$ and $\hat{S}_z$ on $\ket{p}$ produces
\begin{equation}
\begin{aligned}
    	\hat{S}_-\ket{p}&=\hbar\sqrt{(2s-p)(p+1)}\ket{p+1}\\
    	\hat{S}_+\ket{p}&=\hbar\sqrt{p(2s-p+1)}\ket{p-1}\\
    	\hat{S}_z\ket{p}&=\hbar(s-p)\ket{p}.
\end{aligned}    
\end{equation}
The expression in \eqref{coherentState} is equivalent to
\begin{equation}
    \ket{z} \equiv\left(1+|z|^2\right)^{-s}\sum_{p=0}^{2s}\left(\begin{matrix}2s\\p\end{matrix}\right)^{1/2}z^p \ket{p},
\end{equation}
which, as we shall see, is more convenient for computing the action of spin operators on the spin coherent states. In this basis, we can write the partition function \eqref{spinStatesPartitionFunction} as an integral over the complex label $z$ for the spin coherent states as
\begin{equation}
	{\cal Z} = \int d\mu(z)\bra{z}e^{-\beta {\cal \hat{H}}}\ket{z}\label{ZspinCoherentState}
\end{equation}
where the measure must be properly normalized as $\int d\mu(z) \ket{z}\bra{z} = 1$. In this case 
\begin{equation}
    d\mu(z)=\frac{2s+1}{\pi}\frac{dz}{\left(1+|z|^2\right)^2}.
\end{equation}%

\subsection{Crossover from the quantum system to the classical limit}
To study the quantum system close to the classical limit, we must calculate the matrix elements of $\hat{S}_z$ and its powers on the states $\ket{z}$. The first two powers are
\begin{align}
    \bra{z}\hat{S}_z\ket{z} &=\hbar s\frac{1-|z|^2}{1+|z|^2}\\
    \bra{z}\hat{S}^2_z\ket{z} &=\left(\hbar s\frac{1-|z|^2}{1+|z|^2}\right)^2 + 2\hbar^2 s \frac{|z|^2}{(1+|z|^2)^2}.
    \label{secondTerm}
\end{align}
In general, it can be shown that the higher-order terms are all of the form
\begin{equation}
    \bra{z} \hat{S}_z^k \ket{z} =\left(\hbar s\frac{1-|z|^2}{1+|z|^2}\right)^k + \textrm{corrections}.
    \label{eq:sz_powers}
\end{equation}
The first term is the leading term in the classical limit. 
If we were to simply approximate 
\begin{equation}
    \bra{z}\hat{S}_z^k\ket{z}\approx\bra{z}\hat{S}_z\ket{z}^k,
    \label{classLim}
\end{equation}
we would be discarding all quantum fluctuations. However, it is the systematic inclusion of the quantum fluctuations that we aim to achieve in later Section \ref{section_3}.
The second term in \eqref{secondTerm} is an example of a correction term, but there is no general, closed expression for the correction terms of increasing order in $k$. 

These corrections terms expresses the fact that the manifold of the spin states is curved and is not intrinsically due to the noncommutivity of quantum mechanical operators. Essentially these terms are the difference in the trajectory between states on a flat surface compared to a curved surface; rotations on a classical sphere don't commute. However, in taking quantum states to be all possible superpositions of the basis states, the classical states emerge in the limit $\hbar\to 0,s\to\infty$ while keeping the product, ${\sf s}\equiv \hbar s$ fixed. The correction terms are always of the same order in $\hbar$ as the leading term. Thus neglecting these terms does not simply correspond to the semi-classical $\hbar$ expansion and needs to be justified differently. To show this we rewrite equation (\ref{secondTerm}) as
\begin{equation}
 \label{secondTerm1}
 \begin{aligned}
    \bra{z}\hat{S}_z\ket{z} &=\hbar s\frac{1-|z|^2}{1+|z|^2}= {\sf s} \frac{1-|z|^2}{1+|z|^2}\\
    \bra{z}\hat{S}^2_z\ket{z} &=\left(\hbar s\frac{1-|z|^2}{1+|z|^2}\right)^2 + 2\hbar^2 s \frac{|z|^2}{(1+|z|^2)^2}=\\
     & {\sf s}^2\left\{\left(
\frac{1-|z|^2}{1+|z|^2}\right)^2 + 2\frac{1}{s} \frac{|z|^2}{(1+|z|^2)^2}
     \right\},
\end{aligned}
\end{equation}
which highlights the property that the correction terms, which are sensitive to the curvature of the manifold of spin superpositions, are of higher order in an $1/s$ expansion; and that the operators, that have a sensible large-spin, i.e. semi-classical, limit are $\hat{S}_z^k/{\sf s}^k$. Indeed, this limit entails 
taking $\hbar\to 0,s\to\infty$ while keeping the product, ${\sf s}\equiv \hbar s$ fixed. It is precisely these corrections that will be refered to in the rest of the text as noncommutative corrections. Indeed, these corrections arise as the spin coherent states are eigenstates of $\hat{S}^-$ but not of $\hat{S}_z$, and these operators do not commute.

In addition to this double expansion, we are interested in the dependence of the partition function on $\beta$ which characterizes the thermal bath with which our quantum system is in equilibrium. To this end we perform a standard high temperature expansion of the partition function, {\it i.e.} we rewrite the exponential series $e^{-\beta\hat{\cal H}}$ in powers of $\beta$.

Therefore, the corrections to the classical limit we are computing are obtained by a two-fold approximation scheme, both in the noncommutative terms as depicted in \eqref{secondTerm1}, and in the high temperature $\beta$ expansion.

The first term on the right-hand side of \eqref{eq:sz_powers} (ignoring the noncommutative terms), can be written as an exponential series
\begin{equation}
    \sum_{k=0}^\infty\frac{1}{k!}\bra{z} \hat{S}_z^k \ket{z}\approx\exp\left(\hbar s\frac{1-|z|^2}{1+|z|^2}\right), 
    \label{eq:exp_series}
\end{equation}

We now define the Hamiltonian for a single spin (whose electromagnetic properties will be described by its $g-$factor) in an applied magnetic field that is constant along the $z$-direction,
\begin{equation}
    \hat{\cal H}=-\frac{g\muB}{\hbar}\hat{S}_z B_z\label{ZeemanHam}
\end{equation}
For the electron, $g\approx 2.002=|g_e|$ is the absolute value of the electron $g$-factor, $\muB=9.274\times10^{-23}$~J/T is the Bohr magneton, $\hbar=1.05457182\times10^{-34}$~J/K is Planck's constant and $B_z$ is the applied magnetic field in Tesla. Choosing a fixed field direction (which can always be taken to be along $z$) simplifies the calculation by reducing the noncommutativity as we work with the exponential of operators. 

To compute the partition function, we again express the exponential as a series
\begin{equation}
    \exp{\left(-\beta \hat{\cal H}\right)} = \sum_{k=0}^\infty\frac{1}{k!}\left(\beta\frac{g\muB}{\hbar}\hat{S}_zB_z\right)^k,\label{operatorExponentialSeries}
\end{equation}
and compute the matrix elements $\bra{z}\exp(-\beta\hat{\cal H})\ket{z}$, which, using equation \eqref{eq:exp_series}, can be approximated by
\begin{equation}
\begin{aligned}
    \bra{z}\exp(-\beta\hat{\cal H})\ket{z} &\approx \sum_{k=0}^\infty\frac{1}{k!}\left(\beta\frac{g\muB}{\hbar}\right)^k\left(\hbar s\frac{1-|z|^2}{1+|z|^2}\right)^k B^k_z.
\end{aligned}
\end{equation}
Thus, the matrix elements take the simple form
\begin{equation}
    \bra{z}\exp(-\beta\hat{\cal H})\ket{z}\approx\exp\left(\beta g\muB B_z s\frac{1-|z|^2}{1+|z|^2}\right).\label{classicalLimit}
\end{equation}
The complex number $z$ (and its conjugate $\bar{z}$) can then be mapped onto a unit 2-sphere by defining a unit {\it spin coherent state vector}\cite{karchevPathIntegralRepresentation2012}, $\vec{n}$, with components
\begin{equation}
\begin{aligned}
        n_x &= \frac{z+ \bar{z}}{1+|z|^2}\\
        n_y &= -i \frac{ z- \bar{z}}{1+|z|^2}\\
        n_z &= \frac{1-|z|^2}{1+|z|^2},
\end{aligned}
\end{equation}
and using this we can rewrite the matrix elements \eqref{classicalLimit} as
\begin{equation}
    \bra{z}\exp(-\beta\hat{\cal H})\ket{z}\approx\exp\left(\beta g\muB B_z sn_z\right).
\end{equation}
This leads immediately to the definition of the classical Hamiltonian
\begin{equation}
    {\cal H}_\textrm{eff}=-g\muB B_z sn_z=-\mu_s \vec{B}\cdot\vec{S},
    \label{eq:classical_zeeman_hamiltonian}
\end{equation}
where we identify $\vec{S} = \vec{n}$ as the classical spin vector (magnetic moment) with length $\mu_s = sg\muB$. Therefore, dropping the noncommutative terms, indeed yields the expected classical limit of this quantum system. We emphasize that {\em all} the powers of $\hat{S}_z^k$ are needed to recover the classical limit--only the noncommutative terms have been dropped. As we go to the large-spin limit, since the curvature of the sphere is proportional to $1/s^2,$ it becomes smaller and smaller, which justifies neglecting these terms.

The vector $\vec{n}$ defined by the spin coherent states plays the role of the spin unit vector, which is commonly used in classical Heisenberg spin models. Thus, not only does the spin coherent states basis provide us with a continuous (integral) description of the quantum system, but it also yields a straightforward interpretation of the quantum system (described by its states and operators) in terms of the continuous classical system (described by the magnetization vector).
We would like to clarify that the convergence of the quantum infinite spin limit towards the classical limit has been rigorously proven long ago, using spin coherent states, in the more general context of the quantum Heisenberg model in the thermodynamic limit\cite{liebClassicalLimitQuantum1973}, and in more recent work, without using spin coherent states\cite{conlonAsymptoticLimitsQuantum1990} or thermodynamic limit assumptions\cite{millardInfiniteSpinLimit2003}. However, these approaches are based on constructing lower and upper bounds for the quantum partition function (and/or free energy), which converge to the classical limit in the infinite spin limit and don't aim to build a spin dependent classical approximation, which is our goal in this paper. The cornerstone of our model is the double expansion, on the one hand relative to the curvature of the spin manifold--i.e. the {\it $1/s$ expansion} (which can be understood as a {\it large N} or t'Hooft expansion\cite{hooftPlanarDiagramTheory1974}, keeping $\hbar s$ fixed), on the oher hand the high temperature expansion in powers of $\beta$, for the exponential series \eqref{operatorExponentialSeries}. It is from the interplay of these two expansions that we obtain the effective classical Hamiltonian, when equilibrium with both baths (of quantum and thermal fluctuations) is assumed. The aim of our approach is to build an efficient numerical method for computing controlled approximations to the thermal expectation values of the quantum system.

We shall now use the partition function in the spin coherent state basis to compute expectation values for the quantum spin Hamiltonian, close to the classical limit, by performing an expansion in increasing orders of $\beta$. These serve as a reference to compare to numerical calculations using atomistic spin dynamics in Section \ref{section_3}.

\section{Partition function and expectation values\label{section_2}}

The expectation value of an operator $\hat{O}$ for the discrete quantum spin system is
\begin{equation}
    \langle \hat{O} \rangle = \frac{\sum\limits_{m=-s}^{s}\bra{s,m}\hat{O}\exp(-\beta \hat{{\cal H}}) \ket{s,m}}{\sum\limits_{m=-s}^{s}\bra{s,m}\exp(-\beta \hat{{\cal H}}) \ket{s,m}},
\end{equation}
where the denominator is the partition function~\eqref{spinStatesPartitionFunction}. In the spin coherent state basis, the expectation value is expressed in terms of integrals, rather than sums, {\em viz.}
\begin{equation}
    \langle \hat{O} \rangle = \frac{\int d\mu(z) \bra{z}\hat{O}\exp(-\beta \hat{{\cal H}}) \ket{z}}{\int d\mu(z)\bra{z}\exp(-\beta \hat{{\cal H}}) \ket{z}}.
    \label{eq:spin_coherent_expectation}
\end{equation}
As mentioned above, the spin coherent states are not eigenstates of $\hat{S}_z$, making the exponentiation more subtle. The action of the exponential of $\hat{S}_z$ on $\ket{s,m}$ simply yields the exponentiation of the eigenvalue
\begin{equation}
	e^{\hat{S}_z/\hbar}\ket{s,m}=e^{m}\ket{s,m};
\end{equation}
but in the spin coherent state basis, we cannot exactly compute the action and must resort to approximations such as the double expansion described in Section in \ref{section_1} expansion and the high- and low-temperature expansions. 

We proceed by calculating the expectation value $\langle \hat{S}_z\rangle$ as a function of temperature with the Zeeman Hamiltonian~\eqref{ZeemanHam}. This is known to be qualitatively different for classical and quantum spin models due to spin quantisation~\cite{greinerThermodynamicsStatisticalMechanics2000}. The expectation value $\langle \hat{S}_z\rangle$ can be identified with the magnetization induced by an external field (in the limit when the exchange interaction can be neglected).

The exact quantum expectation value, calculated from the discrete basis, where the action of $\hat{S}_z \ket{s,m} = \hbar m \ket{s,m}$, gives
\begin{equation}
    \langle\hat{S}_z\rangle = \frac{\sum\limits_{m=-s}^{s}\hbar m\exp(\beta g\muB m B_z)}{\sum\limits_{m=-s}^{s}\exp(\beta g\muB m B_z)}
    \label{fullyQuantum}.
\end{equation}
The expectation value in the classical limit is calculated with the spin coherent states using equation~\eqref{eq:spin_coherent_expectation} and the approximation in equation~\eqref{classicalLimit} which neglects the terms proportional to powers of $1/s,$ yielding
\begin{equation}
    \langle \hat{S}_z \rangle\approx \hbar s\ddfrac{\int dz \frac{1-|z|^2}{(1+|z|^2)^{3}}\exp\left(\beta g\muB B_z s\frac{1-|z|^2}{1+|z|^2}\right)}{\int dz \frac{1}{(1+|z|^2)^{2}}\exp\left(\beta g\muB B_z s\frac{1-|z|^2}{1+|z|^2}\right)}.\label{CoherentClassicalLimit}
\end{equation}

\begin{figure}
    \centering
    \includegraphics[width=0.48\textwidth]{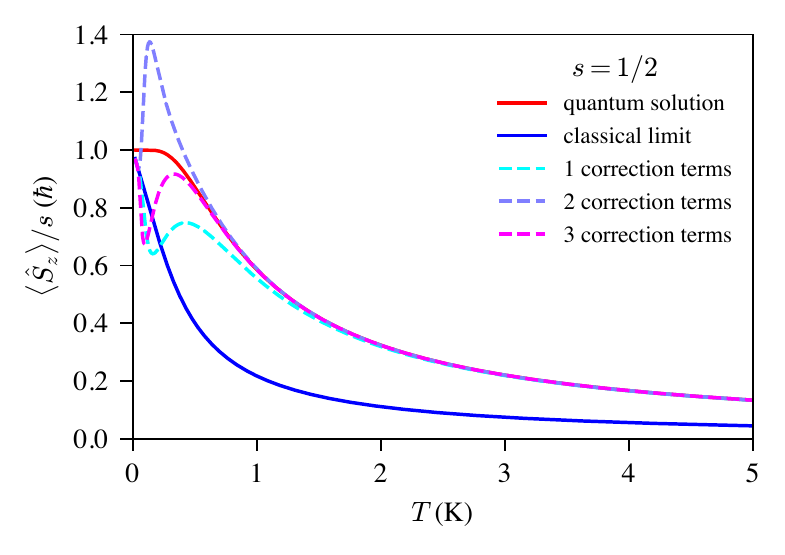}
    \caption{Expectation value $\braket{\hat{S}_z}$ for spin $s=1/2$ as a function of temperature. Red solid line - the exact quantum solution in the discrete spin basis $\ket{s,m}$ from Eq.~\eqref{fullyQuantum}. Blue solid line - the classical limit of the spin coherent state basis from Eq.~\eqref{CoherentClassicalLimit}. Dashed lines are successive corrections to the partition function to include noncommutative terms such as appear in Eq.~\eqref{secondTerm}. The applied field is $B_z = 1$~T for all figures.}\label{PartitionFunctionExpansion}
\end{figure}

Using these expressions for the discrete quantum model \eqref{fullyQuantum} and the classical limit of the spin coherent state \eqref{CoherentClassicalLimit}, we plot the expectation value $\langle \hat{S}_z \rangle$ as a function of temperature in Figure \ref{PartitionFunctionExpansion}. Neglecting the terms due to the non-comutativity of $\hat{S}_z$ and $\hat{S}_\pm,$ i.e. working to leading order in the $1/s$ expansion, means the representation by the spin coherent states produces the classical limit (blue solid line), as expected, with an immediate decay of the spin alignment with the external field as soon as the temperature is non-zero. Equation \eqref{CoherentClassicalLimit} is, in fact, identical to the expectation value $\langle S_z \rangle$ of a classical spin, as is expected from Ehrenfest's theorem--a useful sanity check (see Appendix \ref{app:spin_coherent_classical}). In the quantum case (red solid line) the expectation value remains almost flat--at low temperatures--and displays a slower characteristic decay around the zero temperature value, along with an initial inflection point that is expected on general grounds~\cite{Anders_NewJPhys_24_033020_2022}. 

These characteristic differences between quantum and classical models of single spins are well known and well studied. Of practical interest is that we can obtain an {\em intermediate} approximation for the quantum thermal expectation values by retaining terms related to the noncommutativity of operators. Indeed, if we wish to include quantum features into the classical model in a rigourous manner, we cannot neglect all the noncommutative terms by simply using the approximation of \eqref{classLim}. To do this, the exponential functions in the spin coherent state expectation value \eqref{eq:spin_coherent_expectation} must be expanded as a series in $\beta$,
\begin{equation}
    \exp\left(\beta{\hbar}\hat{S}_z\right)\approx 1+\beta{\hbar}\hat{S}_z+\frac{1}{2}\left(\beta{\hbar}\hat{S}_z\right)^2+\dots\label{second_order exponential}
\end{equation}
Higher-order terms beyond $\hat{S}_z$ contain the effects of the noncommutativity of operators, as seen in \eqref{secondTerm}, and we now include these terms as we evaluate the expectation value. We calculate $\langle\hat{S}_z\rangle$ in the spin coherent state basis in increasing orders of the $\beta$ expansion, which includes the terms due to noncommutativity of $\hat{S}_z$ to higher orders. The results are shown with dashed lines in Figure \ref{PartitionFunctionExpansion}. `1 correction term' includes noncommutative corrections for $\hat{S}_z^2$, `2 correction terms' corrections for $\hat{S}_z^3$ and so on. We see that including even the first noncommuting term in this expansion yields a solution that is already significantly different from the classical result and close to the quantum solution at temperatures of the order of $1$~K and above. The agreement improves as the temperature increases, as expected for an expansion in powers of $\beta$. Going to higher orders in $\beta$ causes the expectation value to converge more quickly to the quantum solution (Figure \ref{PartitionFunctionExpansion}), thus producing a continuous description of the discrete quantum system, which is one of our main objectives.

For {\it very low} temperatures, close to $0$~K, the approximation as a power series in $\beta$ breaks down and diverges because $\beta$ is the inverse of the temperature. We emphasize, however, that already at first order in $\beta$, this semi-classical model accurately captures the salient features of the thermal spin statistics of the quantum system at temperatures of the order of $1$~K. Next, we build a numerical sampling technique for this partition function based on classical, atomistic, spin dynamics.

\section{Effective Hamiltonian and Atomistic spin dynamics\label{section_3}}

\subsection{Low-temperature expansion of the matrix elements}\label{lowTExp}

Building a classical Hamiltonian dynamics model to emulate a quantum system, expressed in the spin coherent states basis, requires finding an effective classical Hamiltonian ${\cal H}_\textrm{eff}$ which approximates $\bra{z}\exp(-\beta \hat{{\cal H}}) \ket{z}$ as $\exp(-\beta{\cal H}_\textrm{eff})$. By finding such an approximate expression, we recast the quantum system with partition function \eqref{spinStatesPartitionFunction} into an effective classical system with partition function
\begin{equation}
\begin{aligned}
        {\cal Z}&=\int d\mu(z)\bra{z}\exp(-\beta \hat{{\cal H}}) \ket{z}\\
        &\approx\int d\tilde{\mu}(z)\exp(-\beta{\cal H}_\textrm{eff}),
        \end{aligned}\label{eq:heff_def}
\end{equation}
where ${\cal H}_\textrm{eff}$ yields the same expectation values as for the quantum case and $\tilde{\mu}(z)$ describes a potentially enlarged, higher-dimensional, phase space, as is the case in path integral molecular dynamics approaches\cite{deymierMultiscaleParadigmsIntegrated2016}.

We consider the partition function with the first noncommutative correction \eqref{secondTerm}, and seek an expression such that
\begin{equation}
\begin{aligned}
\exp\left(-\beta{\cal H}_\textrm{eff}\right) \approx \exp \left(\beta g \muB B_zs\frac{1-|z|^2}{1+|z|^2}\right) \\
+ \frac{1}{s}\left(\beta g \muB B_z s\right)^2 \frac{|z|^2}{\left(1+|z|^2\right)^2},\label{eq:firstApproxH}
\end{aligned}
\end{equation}
where the first term on the right-hand side is the classical limit and the second term is the first noncommutative term which appears on the right-hand side of \eqref{secondTerm}. We ignore higher-order non commutation terms in $\bra{z}\hat{S}_z^k\ket{z}$, beyond $k=2$, keeping only the first noncommutative correction. This is the same level of approximation used in `1 correction term' in Fig.~\ref{PartitionFunctionExpansion}.
As a first and very coarse approximation (for more details, see appendix \ref{app:coarse_approximation}) we take
\begin{equation}
{\cal H}^{\textrm{low-T}}_\textrm{eff}=-g \muB B_zs\frac{1-|z|^2}{1+|z|^2}+ g \muB B_z\frac{\sqrt{2s} |z|}{1+|z|^2},
\label{expApp}
\end{equation}
which, written in terms of the spin coherent state vector $\vec{n}$, is 
\begin{equation}
    {\cal H}^{\textrm{low-T}}_\textrm{eff}=-g \muB B_zs n_z + \tfrac{1}{2}g \muB B_z\sqrt{2s}\sqrt{1-n_z^2}.\label{EffHamn}
\end{equation}
The apparent non analyticity in these equations \eqref{expApp}-\eqref{EffHamn}, is an artifact of our parametrization.

The first term is again the purely classical Zeeman Hamiltonian \eqref{eq:classical_zeeman_hamiltonian}. The second term arises due to the quantization of spin and energetically favors the spin to align with the quantization axis ($z$). It has a form similar to magnetocrystalline anisotropy, but its origin is the quantum behavior of the spin rather than any physical interaction. We will refer to this term as ${\cal H}_\textrm{Qeff}$. 

To calculate the thermal expectation values using this effective Hamiltonian, we use the techniques of atomistic spin dynamics (ASD)~\cite{Halilov_PhysRevB_58_293_1998,Chubykalo_JMagnMagnMater_266_28_2003, Mryasov_EurLett_EPL_69_805_2005,Skubic_JPhysCondensMatter_20_315203_2008,Evans_JPhysCondensMatter_26_103202_2014}. This is usually used to model the dynamics of localized spin magnetic moments $\vec{\mu} = \mu_s\vec{S}$ where $\vec{S}$ is a unit vector and $\mu_s=gs\mu_B$ is the size of the spin magnetic moment. The moments interact with a local effective magnetic field $\vec{B}_\textrm{eff}$ obtained from a Hamiltonian ${\cal H}_\textrm{eff}$ that encodes the different magnetic interactions of the system. Here we will use the normalised vector $\vec{n}$ rather than $\vec{S}$ to emphasize that we are solving the dynamics of the spin coherent state vector rather than making an \textit{a priori} assumption of classical spin magnetic moments. 

Calculations of the thermodynamic quantities of classical spins can be performed with ASD or Monte Carlo calculations, but ASD is trivial to parallelize across large ensembles of spins, allowing efficient calculation as well as the ability to calculate real-time dynamics. The classical spin dynamics is described by the Landau-Lifshitz-Gilbert (LLG) equation of motion
\begin{equation}
	\dot{\vec{n}}=-\frac{\gamma}{1+\alpha^2}\left(\vec{n}\times\vec{B}_\textrm{eff}+\alpha\vec{n}\times\left(\vec{n}\times\vec{B}_\textrm{eff}\right)\right),
    \label{eq:llg}
\end{equation}  
where $\gamma\equiv\frac{g\muB}{\hbar}$ is the gyromagnetic ratio in $\text{rad}\cdot \text{s}^{-1}\cdot\text{T}^{-1}$, $\alpha$ is a dimensionless damping parameter, and the effective field $\vec{B}_\textrm{eff}$ in Tesla is calculated as
\begin{equation}
	\vec{B}_\textrm{eff}=-\frac{1}{\mu_s}\vec{\nabla}_{\vec{n}}{\cal H}\label{EffField}.
\end{equation}
thus, the field from our effective Hamiltonian \eqref{EffHamn} is
\begin{equation}
	\vec{B}^{\textrm{low-T}}_{\textrm{eff}} = B_z\vec{e}_z + \frac{\sqrt{2}}{2\sqrt{s}}B_z\frac{n_z}{\sqrt{n_x^2 + n_y^2}}\vec{e}_z\label{effFieldn},
\end{equation}
where $\vec{e}_z$ is the unit vector along $z$. This expression is apparently singular for $n_z=1;$ this singularity simply indicates that the magnetic field doesn't have any effect on a moment that is aligned with it; we realize, indeed, that such an initial condition, which must be treated separately, is very improbable at any finite temperature. 

Temperature is included in the formalism by adding a stochastic field $\vec{B}_\textrm{eff}\to \vec{B}_\textrm{eff}+\vec{\eta}$ that turns the Landau-Lifshitz-Gilbert equation of motion \eqref{eq:llg} into a Langevin equation. This is where our method gets its path integral name from. We sample the partition function of this system using several stochastic realisations ({\it or paths}) on the Bloch sphere to evaluate the properties of the statistical distribution of the spin vector. The analogue in path integral molecular dynamics methods\cite{Parrinello_JChemPhys_80_860_1984} is using molecular dynamics~\cite{chenBrownianDynamicsMolecular2004} to sample the partition function. The stochastic field is defined through the fluctuation dissipation theorem, which in the classical case requires $\vec{\eta}$ to be a white noise with the properties
\begin{equation}
    \begin{aligned}
    \braket{\eta_i(t)}&=0\\
    \braket{\eta_i(t)\eta_j(t')}&=\frac{2\alpha\delta_{ij}\delta(t-t')}{\beta\mu_s\gamma},
    \end{aligned}
\end{equation}
where $i,j$ are Cartesian components. 

In our work the quantum nature of the spin is included directly into the effective field without making any assumption of the statistical distribution.

Recently, stochastic fields using the quantum fluctuation dissipation theorem have been used, enforcing a Bose-Einstein statistical distribution for the noise \cite{Barker_PhysRevB_100_140401_2019}. This assumes that the relevant thermally occupied objects in this case are magnons, which should obey bosonic statistics.

We numerically integrate the LLG equation \eqref{eq:llg} using a symplectic integration scheme \cite{Thibaudeau_PhysAStatMechitsAppl_391_1963_2012} with a timestep of $0.05$~ps. The expectation values from the numerical method are calculated as averages over time and multiple realizations of the stochastic dynamics
\begin{equation}
	\braket{S_z} =\frac{1}{N_s}\frac{1}{N_t} \sum_{i=1}^{N_s} \sum_{t=1}^{N_t} n_{i,z}(t),\label{atomisticAverage}
\end{equation}
where $N_s$ is the number of independent spin trajectories and $N_t$ is the number of time samples. The average in time is taken after an equilibration period where the system relaxes from the initial state to a thermalized state. The simulations performed here equilibrate within a few nanoseconds; therefore, we started the averaging procedure after an equilibration period of $5$~ns. The averaging time is $15$~ns and $N_s = 20$.

\begin{figure}
	\centering
	\includegraphics[width=0.5\textwidth]{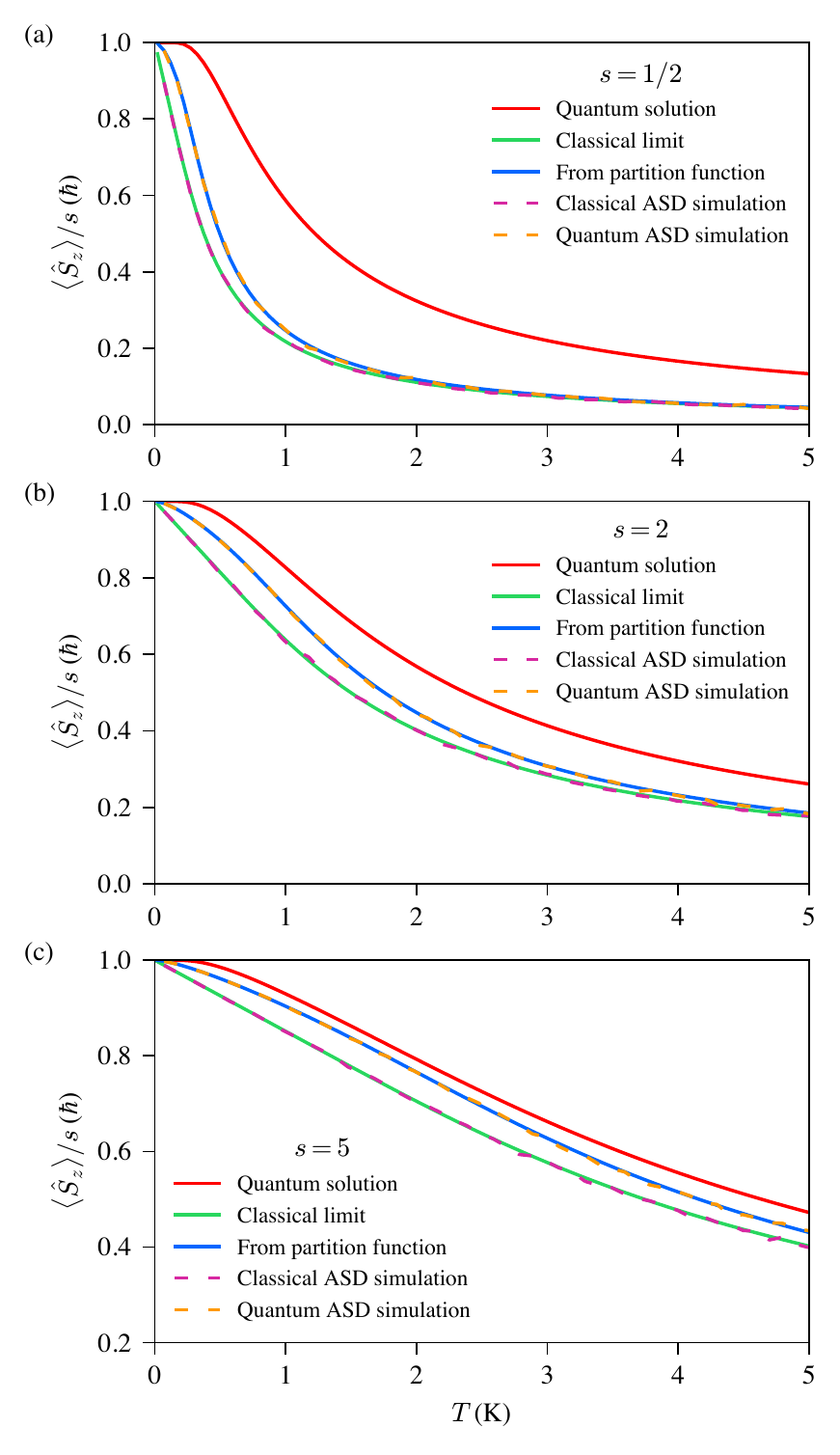}
	\caption{Expectation value for $\hat{S}_z$ as a function of temperature for the classical limit (green solid curve), quantum solution (red solid curve) and effective model (blue solid curve) from partition function. Equivalent results from enhanced atomistic spin dynamics simulation for classical limit (purple dashed curve) and effective model (orange dashed curve). (a) Top pane $s=1/2$, (b) middle pane $s=2$ and (c) bottom pane $s=5$}\label{qasds}
\end{figure}

From the effective Hamiltonian \eqref{expApp}, we compute the expectation value for $\hat{S}_z$
\begin{equation}
	\braket{\hat{S}_z}\approx\frac{\int d\mu(z)\hbar s\frac{1-|z|^2}{1+|z|^2}\exp(-\beta {\cal H}_\textrm{eff})}{\int d\mu(z)\exp(-\beta {\cal H}_\textrm{eff})},\label{resFromParFunction}
\end{equation}
and compare these to results from \eqref{atomisticAverage}.
The results for different values of the principal quantum number $s=1/2,~2,~5$ are shown in Figure \ref{qasds}.

All three models, classical, quantum and the effective Hamiltonian \eqref{resFromParFunction} converge to the same values in the high-temperature limit. In figure~\ref{qasds}a for $s=1/2$ the effective model differs only slightly from classical model and is not close to the quantum model. Only the slope at zero temperature shows any of the quantum behavior with a small inflection point. This is a feature which several effective models have attempted to force artificially on the studied spin systems to reproduce the experimental behavior for magnetization curves~\cite{Kuzmin_PhysRevLett_94_107204_2005}. However, our effective classical atomistic model does not impose any assumptions on the system and has no fitting parameters. The additional computational cost of making the classical system more closely resemble its quantum avatar is minimal, requiring only the addition of a field that amounts to an effective anisotropy.

Although this coarse approximation scheme provides results that are closer to the quantum results, there is no way to systematically improve it. For each higher-order noncommutative correction we must again try to derive a $\mathcal{H}_{\mathrm{eff}}$ ad hoc that satisfies equation~\eqref{eq:heff_def}. Therefore, we continue by developing a more systematic method for which computing the thermal expectation values to higher orders of accuracy is straightforward.

\subsection{High-temperature spin coherent states expansion}\label{high-t-model}

The effective model in the previous section produced by approximating the integrand of the partition function by an exponential is very coarse but yields part of the quantum corrections and at a very low computational cost. We now improve on this to try to recover a behavior more similar to the expansion of the partition function in Figure~\ref{PartitionFunctionExpansion}. We do this by including higher-order noncommutative terms in the expansion of $\exp(-\beta\hat{\mathcal{H}})$ \eqref{operatorExponentialSeries} in a more systematic way.

To this end, we return to the partition function \eqref{ZspinCoherentState} and, similar to the path-integral molecular dynamics approaches, introduce the resolution of unity as
\begin{equation}
	\sum_{p=0}^{2s}\ket{p}\bra{p}=1,
\end{equation}
in the $\ket{s,m}$ basis, in which $\hat{S}_z$ is diagonal, resulting in
\begin{equation}
	{\cal Z}=\int \sum_{p=0}^{2s}d\mu(z)\bra{z}e^{\frac{\beta g\mu_B}{\hbar}B_z\hat{S}_z}\ket{p}\braket{p|z}.
\end{equation}
Using the definition of $\ket{z}$ and the action of $\hat{S}_z$ on $\ket{p}$ we find
\begin{equation}
	{\cal Z}=\int d\mu(z)\left[e^{-\beta g\muB s B_z}\left(\frac{e^{\beta g\mu_B B_z}+|z|^2}{1+|z|^2}\right)^{2s}\right],\label{Coherent}
\end{equation}
for which we need to rewrite the integrand
\begin{equation}
	F[\beta, z]\equiv e^{-\beta g\muB s B_z}\left(\frac{e^{\beta g \mu_B B_z}+|z|^2}{1+|z|^2}\right)^{2s},\label{toMakeExponential}
\end{equation}
as a single exponential of the form $F[\beta, z]\equiv \exp({-\beta{\cal H}_\textrm{eff}})$ in order to identify an effective Hamiltonian. Through a series of identities (see appendix \ref{app:HighT_exponential}), we can write 
\begin{equation}
    \begin{aligned}
F[\beta, z] &=
	\exp\left\{2s\left[\ln(2)+\ln\left(\frac{|z|}{1+|z|^2}\right) \right.\right.\\
 &\left.\left.+\ln\left(\cosh\left(e^{\frac{\beta g\mu_B B_z}{2}}-\ln\left(|z|\right)\right)\right)\right]\right\}.\label{eq:FasExp}
     \end{aligned}
\end{equation}
At this stage, the expression is still exact and includes all noncommutative corrections to the classical limit and all orders of temperature. We then approximate \eqref{eq:FasExp} with a Taylor expansion as $\beta \to 0$. Thus in the high-temperature limit (which we later find to be quite low)

\begin{equation}
\begin{aligned}
\ln(F[\beta, z])&\approx\frac{\left(1-|z|^2\right) \beta g\muB s B_z}{1+|z|^2}+\frac{|z|^2 \beta^2 \left(g\mu_B\right)^2 sB_z^2}{\left(1+|z|^2\right)^2}\\
                &-\frac{|z|^2 \left(1-|z|^2\right) \beta^3\left(g\mu_B\right)^3 s B_z^3}{3 \left(1+|z|^2\right)^3} + \mathcal{O}(\beta^4).
\label{expApprox}
\end{aligned}
\end{equation}
Mapping to the spin coherent state vector components using $(1-|z|^2)/(1+|z|^2)=n_z$ and $|z|^2/(1+|z|^2)=(1-n_z^2)/4$, we can write a temperature-dependent effective Hamiltonian:
\begin{equation}
\begin{aligned}
{\cal H}^{\textrm{high-T}}_\textrm{eff}\approx &- g\muB s B_z n_z - \tfrac{1}{4}\beta \left(g\mu_B\right)^2 sB_z^2(1-n_z^2)\\
    &+\tfrac{1}{12}\beta^2\left(g\mu_B\right)^3 s B_z^3 n_z(1-n_z^2).
    \label{lowTeffHam}
\end{aligned}
\end{equation}
From the temperature-dependent Hamiltonian \eqref{lowTeffHam} and the definition of the effective field \eqref{EffField}, we derive
\begin{equation}
	\vec{B}^{\textrm{high-T}}_\textrm{eff}=B_z-\tfrac{1}{2}\beta g \mu_B B_z^2 n_z-\tfrac{1}{12}\beta^2\left(g\mu_B\right)^2B_z^3(1-3n_z^2).
	\label{lowTeffField}
\end{equation}

\begin{figure}
	\centering
	\includegraphics[width=0.5\textwidth]{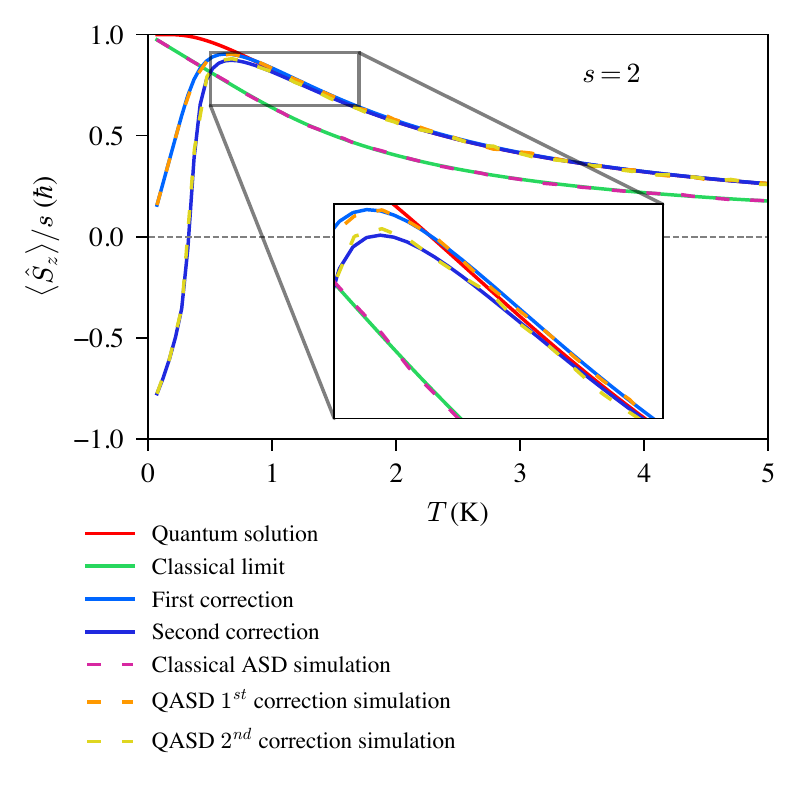}
	\caption{Expectation value for $\hat{S}_z$ for $s=2$ as a function of temperature for classical limit (green solid curve) and quantum solution (red solid curve) and effective model with the first correction (light blue solid curve) and second correction (dark blue solid curve) from partition function. Equivalent results from enhanced atomistic spin dynamics simulation for classical limit (purple dashed curve) and second effective model with first correction (orange dashed curve) and second correction (yellow dashed curve)}\label{qasds2}
\end{figure}

We use this effective field in numerical atomistic simulations, and sample several stochastic paths of these effective dynamics over the Bloch sphere. We compare the results with the expectation values computed directly from the partition function \eqref{resFromParFunction} and the relevant terms, according to the order of the approximation, of the effective Hamiltonian \eqref{lowTeffHam}. The results are shown in Figure \ref{qasds2}. 

When we include only the first correction for the effective field, namely the first and second terms on the right-hand side of \eqref{lowTeffHam} then, contrary to the previous section (Figure \ref{qasds}), the low-temperature limit is far from both classical and quantum solutions. However, around $1$~K, the results become very close to the quantum solution and converge to be almost identical as the temperature increases.

Including higher-order terms (for example, using all the terms in \eqref{lowTeffField}) we see that although at low temperatures the model is initially further away from the quantum solution, the rate of convergence towards the quantum model is much faster than for lower order corrections. For the first correction, once close to the quantum solution, it takes a while before both curves are indistinguishable, and this happens much quicker when including the second term (see the inset of Figure \ref{qasds2}). As our approximation is computed to higher orders, the convergence becomes faster. We note that there is no reason why this high-temperature expansion should become valid at much lower temperatures as we go to higher orders. 

Another issue that we have to deal with is that these expectation values have to be normalized in order for the atomistic simulations to overlap with the direct computation from the partition function. 
Indeed, when we compute the expectation value for $\braket{\hat{S}_z}$ we should be using an expression of the form of Eq.~\eqref{eq:spin_coherent_expectation} as
\begin{equation}
    \langle \hat{S}_z \rangle\approx \frac{\int d\mu(z)\bra{z}\hat{S}_z\exp\left(\frac{\beta g\mu_B}{\hbar}B_z\hat{S}_z\right)\ket{z}}{\int d\mu(z)e^{-\beta\mu_sB_z}\left(\frac{e^{\beta g\mu_BB_z}+|z|^2}{1+|z|^2}\right)^{2s}},\label{eq:partFunction}
\end{equation}
but instead (see appendix \ref{app:HighT_normalisation}), the consistent approximation is given by
\begin{equation}
    \langle \hat{S}_z \rangle_\textrm{app}\equiv \frac{\int d\mu(z)\hbar s\frac{1-|z|^2}{1+|z|^2}e^{-\beta\mu_sB_z}\left(\frac{e^{\beta g \mu_BB_z}+|z|^2}{1+|z|^2}\right)^{2s}}{\int d\mu(z)e^{-\beta\mu_sB_z}\left(\frac{e^{\beta g\mu_BB_z}+|z|^2}{1+|z|^2}\right)^{2s}}.\label{SecondApprox}
\end{equation}
We know that in the quantum case given by Eq.~\eqref{fullyQuantum}, $\braket{\hat{S_z}}_\textrm{quantum}$ goes to $s$ as $\beta\to\infty$. We can show that in the same limit, for Eq.~\eqref{eq:appExpVal2}, we have
\begin{equation}
	\braket{\hat{S}_z}_\textrm{app}\xrightarrow[\beta\to\infty]{}\frac{s^2}{s+1}\label{normalisationFactor}
\end{equation}
hence our expectation values need to be normalized by this factor to yield the correct results (see appendix \ref{app:HighT_normalisation} for more details).

In summary, using this approximation scheme, we can compute expectation values for the quantum system from an equivalent classical atomistic simulation where the quantum nature of the system is represented by a {\it temperature-dependent} effective field. This is not a surprise as the space of states is curved. In contrast to the previous section (\ref{lowTExp}), these then need to be properly rescaled. However, we can compute a closed expression for this rescaling factor, which once again depends only on the principal quantum spin number $s$. Once this step is fulfilled, the results are almost identical to the fully quantum expectation values for high enough temperatures, which are of the order of $1$~K for the single spin in a magnetic field studied here. The low-temperature behavior of this scheme is not as well behaved as in Section \ref{lowTExp}, which is not surprising, as this is a high-temperature expansion (see Appendix \ref{app:high-t-10th-order}).

\section{Conclusions and outlook}\label{conclusion}

In this Article, we have built an effective, classical, dynamical model for quantum spin systems from a path integral approach inspired by path integral molecular dynamics in the simplest case of a single spin of arbitrary size in a constant magnetic field described by a Zeeman Hamiltonian. While path integral models of spin have a long history and have been investigated in fundamental contexts such as supersymmetry or, more closely related to our work for molecular magnets, a systematic approach bridging the gap from small-size fully quantum simulations to large-scale dynamical simulations with quantum features has been lacking. Our work here is the first step towards this direction.

We have started by expressing the partition function for spin systems in the spin coherent state basis to obtain a continuous description in terms of an integral rather than a sum, to make the connection to classical spin dynamics. This allows the use of highly efficient atomistic spin dynamics simulations for quantum spin systems and makes the connection between the quantum system defined by its states and the Hamiltonian operator and classical spin dynamics more explicit. We then proceeded to expand the relevant matrix elements of the partition function in powers of $\beta$ to compute the expectation values of $\hat{S}_z$ directly from the partition function and from atomistic spin dynamics. Here, we have seen that in this first approximation this could be done very simply and efficiently by adding an anisotropic effective field, which could be directly inferred from the quantum spin number of the system. For small spin values, we have seen that the improvement is quite small but increases with the spin. Of course, spin $s=1/2$ represents the most extreme limit of spin quantization. As the magnitude of the spin increases to $s=2$ and $s=5$ (Fig.~\ref{qasds}b,c) the corrections in the effective model take the system closer to the quantum solution. Many magnetic materials of practical relevance have $s$ in the range $3/2$ to $7/2$ so having an improved quantum description for these larger spin values is already very useful.

We also investigated a different method of approximating the integrand of the partition function by an exponential by allowing the effective Hamiltonian of the system to be explicitly temperature-dependent, yielding a temperature-dependent effective field for describing in this way the quantum nature of the system. This method proved to be more accurate for higher temperatures, above $1$~K, than the low-temperature expansion, but with the drawback that the expectation values computed using this method require renormalization. However, this renormalization factor has a closed general expression that depends only on the quantum spin number $s$ of the system.

The next step we aim to investigate is the more general case of a general, time-dependent, magnetic field. This introduces more noncommutativity issues with operators $\hat{S}_x$, $\hat{S}_y$ and $\hat{S}_\pm$. Beyond this, more complex Hamiltonians including the exchange interaction and magnetocrystalline anisotropy in a quantum fashion will allow the large-scale calculation of the thermodyamics of magnetic materials including quantum effects with a relatively low computational cost. In the present case of a constant magnetic field and for a single spin, we have seen that, conversely to path integral methods for molecular dynamics, we did not need to introduce copies of the spin which interact with itself. We do not expect this to hold in more complex Hamiltonians.

It is important to note that despite being a dynamical sampling method, our method provides accurate results for the evalutation of the thermal expectation values, but is not guaranteed to provide accurate real-time dynamics when quantum fluctations drive the system {\it far} from the classical limit. In future studies we aim to explore how the dynamical behaviour changes in this context and we expect some fundamental aspects of spin path integrals\cite{zwanzigerBerryPhase1990} which apparently do not arise in our model, to resurface for real-time dynamics, even at higher temperatures.

\section*{Data Access}
Python code and output data to reproduce all results and figures reported in this paper are openly available from the Zenodo repository: \textit{Sources for: Numerical Simulations of a Spin Dynamics Model Based on a Path Integral Approach.} \url{https://doi.org/10.5281/zenodo.7692092}~\cite{thisDataset}. The repository contains:
\begin{itemize}
    \item Python code to generate analytic equations derived herein.
    \item Python code to perform enhanced atomistic spin dynamics calculations with the quantum effective fields.
    \item Python scripts to reproduce all figures.
\end{itemize}
The software and data are available under the terms of the MIT License.

\section*{Author Contributions}

Thomas Nussle: conceptualization, methodology, investigation, software, writing - original draft. Stam Nicolis: methodology, writing - review and editing. Joseph Barker: conceptualization, methodology, software, data curation, writing - review and editing, funding acquisition.

\section*{acknowledgments}

This work was supported by the Engineering and Physical Sciences Research Council [grant number EP/V037935/1]. JB acknowledges funding from a Royal Society University Research Fellowship. The authors thank A. Sylla, F. Lab\'{e}y and T. Raujouan for very insightful mathematical discussions, as well as J. Hodrien and A. Coleman from the University of Leeds Research Computing team for their help with optimizing the Python code on which this work is relying. 

\appendix

\section{Correspondence of the spin coherent states with the classical limit}
\label{app:spin_coherent_classical}
Here we show that the observable $\braket{\hat{S}_z}$ from the spin coherent states with the commutators neglected (i.e. in the classical limit \eqref{CoherentClassicalLimit}) is identical to $\braket{S_z}$ calculated from the classical Heisenberg model. For a classical Heisenberg spin with Hamiltonian
\begin{equation}
    {\cal H}=-\mu_s\vec{B}\cdot\vec{S},
\end{equation}
where $\vec{S}$ lives on the unit sphere, the partition function is
\begin{equation}
    {\cal Z} =\int d\vec{S}\delta(\vec{S}^2-1)e^{-\beta{\cal H}}=\int d \vec{S}\delta(\vec{S}^2-1)e^{\beta\mu_s\vec{B}\cdot\vec{S}},
\end{equation}
for which the expectation value of the $z$-component of $\vec{S}$ is given by
\begin{equation}
    \braket{S_z}=\frac{\displaystyle\int d\vec{S}\delta(\vec{S}^2-1)S_z e^{\beta\mu_s\vec{B}\cdot\vec{S}}}{\displaystyle\int d\vec{S}\delta(\vec{S}^2-1)e^{\beta\mu_s\vec{B}\cdot\vec{S}}}.
\end{equation}
If the external field is constant along the $z$-direction then we have
\begin{equation}
    \braket{S_z}=\frac{\displaystyle\int dS_zS_z e^{\beta\mu_sB_z S_z}}{\displaystyle\int dS_ze^{\beta\mu_sB_z S_z}}\label{eq:ClassicalHeisenbergPartition}
\end{equation}
as the integrals over $S_x$ and $S_y$ in the numerator and denominator cancel each other out. Comparing this to $\braket{\hat{S}_z}$ for the spin coherent state \eqref{CoherentClassicalLimit} and using $n_z=(1-|z|^2)/(1+|z|^2)$ and $\mu_s S_z = gs\muB n_z$ we see that \eqref{eq:ClassicalHeisenbergPartition} and \eqref{CoherentClassicalLimit} are identical up to a factor of $\hbar$, as the classical spin vector has no units, whereas the quantum expectation value of $\braket{\hat{S}_z}$ is in units of $\hbar$.

\section{Coarse approximation method}
\label{app:coarse_approximation}

We expand the operator exponential series \eqref{operatorExponentialSeries} up to second order in $\beta$
\begin{equation}
\begin{aligned}
    \exp(-\beta \hat{{\cal H}}) &\approx 1+\beta g \muB B_zs\frac{1-|z|^2}{1+|z|^2}\\
    &+\beta^2 \left(g \muB B_z\right)^2\frac{s |z|^2}{\left(1+|z|^2\right)^2}\\
    &+\frac{1}{2}\left(\beta g \muB B_zs\frac{1-|z|^2}{1+|z|^2}\right)^2,\label{secondOrderOperatorSeries}
\end{aligned}
\end{equation}
we can show that by taking
\begin{equation}
    {\cal H}_\textrm{eff}=-g \muB B_zs\frac{1-|z|^2}{1+|z|^2}+ g \muB B_z\frac{\sqrt{2s} |z|}{1+|z|^2},\label{eq:expApproxCoarse}
\end{equation}
and expanding the effective classical exponential up to the same order in $\beta$, we get
\begin{equation}
\begin{aligned}
    &\exp(-\beta {\cal H}_\textrm{eff})\\
    &\approx 1+\beta g \muB B_zs\frac{1-|z|^2}{1+|z|^2}+\beta^2 \left(g \muB B_z\right)^2\frac{s |z|^2}{\left(1+|z|^2\right)^2}\\
    &+\frac{1}{2}\left(\beta g \muB B_zs\frac{1-|z|^2}{1+|z|^2}\right)^2\\
    &- \beta g \muB B_z\frac{\sqrt{2s}|z|}{1+|z|^2}-\left(\beta g \muB B_z\right)^2\frac{s\sqrt{2s}|z|\left(1-|z|^2\right)}{\left(1+|z|^2\right)^2}.\label{eq:coarseApprox}
\end{aligned}
\end{equation}
This is where our approximation becomes more qualitative than quantitative. Indeed, the fifth and sixth terms on the right-hand side of \eqref{eq:coarseApprox} are not present in \eqref{secondOrderOperatorSeries} even though they are not of higher order in $\beta$, however, we have taken advantage of the freedom of choice for the sign of the extra term in the effective Hamiltonian (second term on the right-hand side of \eqref{eq:expApproxCoarse}) as the correction (third term on the right-hand side of \eqref{secondOrderOperatorSeries}) comes from the square term in the exponential series. Taking the correction (second term on the right-hand side of \eqref{eq:expApproxCoarse}) to be negative implies that
\begin{equation}
	\exp\left(-\beta g \muB B_z\frac{\sqrt{2s} |z|}{1+|z|^2}\right) \in[0;1],
\end{equation}
or in terms of the spin coherent state vector
\begin{equation}
	\exp\left(-\beta \tfrac{1}{2}g \muB B_z\sqrt{2s}\sqrt{1-n_z^2}\right) \in[0;1],
\end{equation}
which means that our expectation value remains close to the classical expectation value, especially for lower temperatures where the spin preferentially aligns with the $z$-axis. Although this constitutes quite a coarse approximation, it is definitely a relevant primer to understand the subtleties of the path integral spin dynamics method.

\section{High temperature model exponential form}
\label{app:HighT_exponential}

Starting from \eqref{toMakeExponential}, we rewrite 
\begin{equation}
\begin{aligned}
	&\left(\frac{e^{\beta g \mu_B B_z}+|z|^2}{1+|z|^2}\right)^{2s}= \left(\frac{e^{\beta g \mu_B B_z}+e^{2\ln(|z|)}}{e^{\ln(1+|z|^2)}}\right)^{2s}\\
	&=\left(\frac{e^{\frac{\beta g \mu_B B_z}{2} + \ln(|z|)}\left(e^{\frac{\beta g \mu_B B_z}{2} - \ln(|z|)}+e^{-\frac{\beta g \mu_B B_z}{2} + \ln(|z|)}\right)}{e^{\ln(1+|z|^2)}}\right)^{2s}\\
	&=\left(\frac{e^{\frac{\beta g \mu_B B_z}{2} + \ln(|z|)}2\cosh\left(\frac{\beta g \mu_B B_z}{2} - \ln(|z|)\right)}{e^{\ln(1+|z|^2)}}\right)^{2s}\\
	&=\left(e^{\frac{\beta g \mu_B B_z}{2} + \ln(\frac{|z|}{1+|z|^2})+\ln\left(2\cosh\left(\frac{\beta g \mu_B B_z}{2} - \ln(|z|)\right)\right)}\right)^{2s},
 \end{aligned}
\end{equation}
hence \eqref{toMakeExponential} can be rewritten as
\begin{equation}
	F[\beta, z]=e^{2s\left(\ln(2)+\ln(\frac{|z|}{1+|z|^2})+\ln\left(\cosh\left(\frac{\beta g \mu_B B_z}{2} - \ln(|z|)\right)\right)\right)}.
\end{equation}

\section{High-temperature model normalization}
\label{app:HighT_normalisation}

We approximate
\begin{equation}
	\begin{aligned}
	&\bra{z}\hat{S}_z\exp\left(\frac{\beta\mu_s}{\hbar}B_z\hat{S}_z\right)\ket{z}\\
	&\approx\bra{z}\hat{S}_z\ket{z}\bra{z}\exp\left(\frac{\beta\mu_s}{\hbar}B_z\hat{S}_z\right)\ket{z}\\
	&=\hbar s\frac{1-|z|^2}{1+|z|^2}e^{-\beta\mu_sB_zs}\left(\frac{e^{\beta\mu_sB_z}+|z|^2}{1+|z|^2}\right)^{2s},
\end{aligned}
\end{equation}
as our approximation scheme for the partition function aims to move from a quantum description in terms of states and operators to a classical description
\begin{equation}
	\bra{z}\exp\left(\frac{\beta\mu_s}{\hbar}B_z\hat{S}_z\right)\ket{z}\approx \exp\left(-\beta{\cal H}\right).
\end{equation}
Within this approximation, we can rewrite
\begin{equation}
\begin{aligned}
    &\frac{\int d\mu(z)\bra{z}\hat{S}_z\exp\left(\frac{\beta\mu_s}{\hbar}B_z\hat{S}_z\right)\ket{z}}{\int d\mu(z)e^{-\beta\mu_sB_zs}\left(\frac{e^{\beta\mu_sB_z}+|z|^2}{1+|z|^2}\right)^{2s}} \\
    &\equiv \frac{\int d\mu(z)\hbar s\frac{1-|z|^2}{1+|z|^2}e^{-\beta\mu_sB_zs}\left(\frac{e^{\beta\mu_sB_z}+|z|^2}{1+|z|^2}\right)^{2s}}{\int d\mu(z)e^{-\beta\mu_sB_zs}\left(\frac{e^{\beta\mu_sB_z}+|z|^2}{1+|z|^2}\right)^{2s}},\label{eq:appExpVal2}
    \end{aligned}
\end{equation}
which is the expression we use for our averages, as it corresponds to the same approximation as the atomistic model, as proven by the exact overlap of both the averages computed from the partition function \eqref{SecondApprox} and the atomistic average over time and the number of realizations \eqref{atomisticAverage}.

What is of peculiar interest is that the ratio
 \begin{equation}
 	\frac{\braket{\hat{S}_z}_\textrm{app}}{\braket{\hat{S}_z}_\textrm{quantum}}\xrightarrow[\beta\to\infty]{}\frac{s}{s+1}
 \end{equation}
which reminds us of the fact that the  eigenvalues of $\hat{\vec{S}}^2$ are $s(s+1)$ as in
\begin{equation}
	\hat{\vec{S}}^2\ket{s,m}=s(s+1)\ket{s,m}
\end{equation}
rather than simply $s^2$. Indeed, in the classical limit $s\to\infty$ we recover
\begin{equation}
	s(s+1)\xrightarrow[s\to\infty]{}s^2.
\end{equation}
 
We would like to emphasize that this required normalization factor is identical for both the results of the atomistic simulations \eqref{atomisticAverage} and the results from the approximate partition function \eqref{SecondApprox}.
 
\begin{figure}
	\centering
	\includegraphics[width=0.5\textwidth]{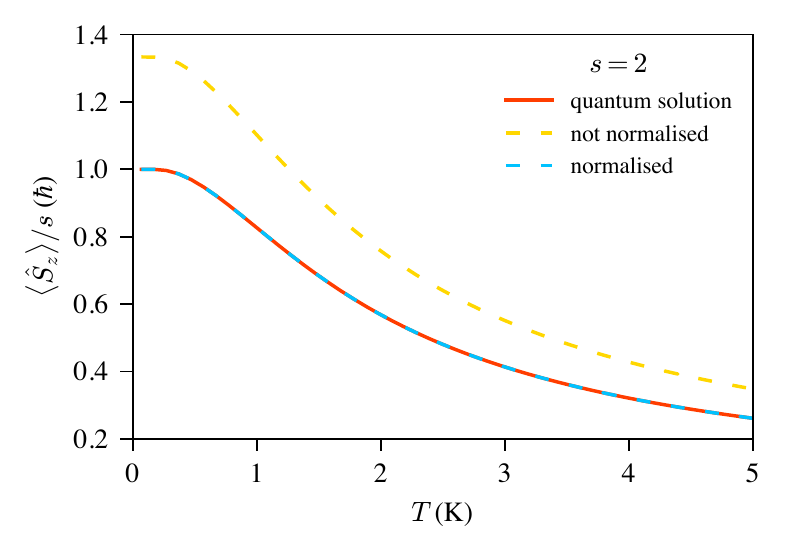}
	\caption{Expectation value for $\hat{S}_z$ for $s=2$ as a function of temperature from \eqref{SecondApprox} (orange dashed curve) and normalised according to \eqref{normalisationFactor} (cyan dashed curve) compared to the quantum limit (red solid curve)}\label{normalisationPartition}
\end{figure}

The expectation values for $\braket{\hat{S}_z}_\textrm{app}$ with and without normalization are given in Figure \ref{normalisationPartition}, along with the appropriate quantum solution.

This is very important for more general applications of this model as this means that the normalization of the curves does not require an additional fitting parameter of any kind but is rather analytically computable and has a general, closed expression.

\section{Higher order correction for the high-temperature model}\label{app:high-t-10th-order}

\begin{figure}
	\centering
	\includegraphics[width=0.5\textwidth]{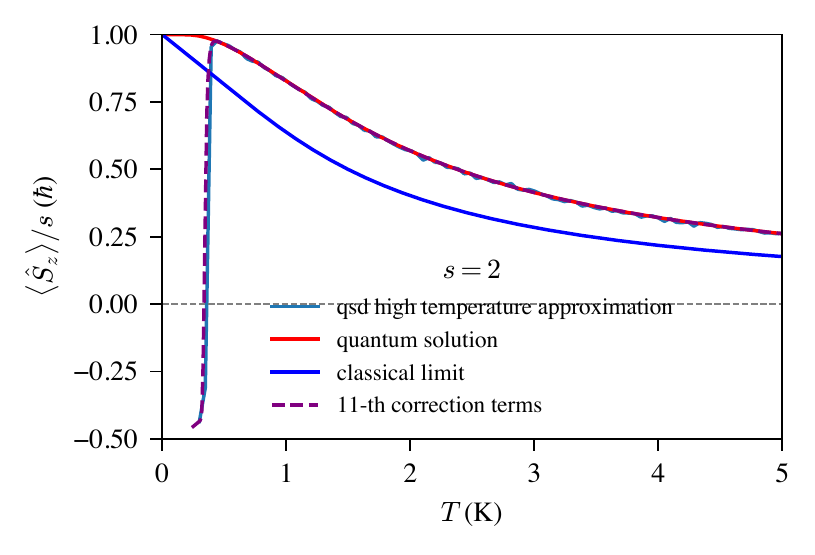}
	\caption{Expectation value for $\hat{S}_z$ for $s=2$ as a function of temperature for classical limit (blue solid curve) and quantum solution (red solid curve) and effective model with the $10^\textrm{th}$ correction (light blue solid curve) from partition function. Equivalent results from enhanced atomistic spin dynamics simulation effective model with $11^\textrm{th}$ correction (purple dashed curve).}\label{high-order-correction}
\end{figure}

As mentioned in section \ref{high-t-model} our method can technically carry out this approximation scheme to any order in the noncommutative terms, numerically, without requiring to compute these corrections using pen and paper. But as this relies on a Taylor expansion around the high-temperature limit $\beta\to0$ there is a limit as to how low in temperature we can provide accurate results. Indeed there is no reason for this high-temperature expansion to converge to the quantum solution for temperatures around $0$~K. This is shown in Figure~\ref{high-order-correction}.

\bibliography{bibliography}

\begin{thebibliography}{43}%
\makeatletter
\providecommand \@ifxundefined [1]{%
 \@ifx{#1\undefined}
}%
\providecommand \@ifnum [1]{%
 \ifnum #1\expandafter \@firstoftwo
 \else \expandafter \@secondoftwo
 \fi
}%
\providecommand \@ifx [1]{%
 \ifx #1\expandafter \@firstoftwo
 \else \expandafter \@secondoftwo
 \fi
}%
\providecommand \natexlab [1]{#1}%
\providecommand \enquote  [1]{``#1''}%
\providecommand \bibnamefont  [1]{#1}%
\providecommand \bibfnamefont [1]{#1}%
\providecommand \citenamefont [1]{#1}%
\providecommand \href@noop [0]{\@secondoftwo}%
\providecommand \href [0]{\begingroup \@sanitize@url \@href}%
\providecommand \@href[1]{\@@startlink{#1}\@@href}%
\providecommand \@@href[1]{\endgroup#1\@@endlink}%
\providecommand \@sanitize@url [0]{\catcode `\\12\catcode `\$12\catcode
  `\&12\catcode `\#12\catcode `\^12\catcode `\_12\catcode `\%12\relax}%
\providecommand \@@startlink[1]{}%
\providecommand \@@endlink[0]{}%
\providecommand \url  [0]{\begingroup\@sanitize@url \@url }%
\providecommand \@url [1]{\endgroup\@href {#1}{\urlprefix }}%
\providecommand \urlprefix  [0]{URL }%
\providecommand \Eprint [0]{\href }%
\providecommand \doibase [0]{http://dx.doi.org/}%
\providecommand \selectlanguage [0]{\@gobble}%
\providecommand \bibinfo  [0]{\@secondoftwo}%
\providecommand \bibfield  [0]{\@secondoftwo}%
\providecommand \translation [1]{[#1]}%
\providecommand \BibitemOpen [0]{}%
\providecommand \bibitemStop [0]{}%
\providecommand \bibitemNoStop [0]{.\EOS\space}%
\providecommand \EOS [0]{\spacefactor3000\relax}%
\providecommand \BibitemShut  [1]{\csname bibitem#1\endcsname}%
\let\auto@bib@innerbib\@empty
\bibitem [{\citenamefont {Ceperley}\ and\ \citenamefont
  {Alder}(1986)}]{Ceperley_Science_231_555_1986}%
  \BibitemOpen
  \bibfield  {author} {\bibinfo {author} {\bibfnamefont {David}\ \bibnamefont
  {Ceperley}}\ and\ \bibinfo {author} {\bibfnamefont {Berni}\ \bibnamefont
  {Alder}},\ }\bibfield  {title} {\enquote {\bibinfo {title} {{Quantum} {Monte}
  {Carlo}},}\ }\href {\doibase 10.1126/science.231.4738.555} {\bibfield
  {journal} {\bibinfo  {journal} {Science}\ }\textbf {\bibinfo {volume}
  {231}},\ \bibinfo {pages} {555--560} (\bibinfo {year} {1986})}\BibitemShut
  {NoStop}%
\bibitem [{\citenamefont {Barker}\ and\ \citenamefont
  {Bauer}(2019)}]{Barker_PhysRevB_100_140401_2019}%
  \BibitemOpen
  \bibfield  {author} {\bibinfo {author} {\bibfnamefont {Joseph}\ \bibnamefont
  {Barker}}\ and\ \bibinfo {author} {\bibfnamefont {Gerrit E.~W.}\ \bibnamefont
  {Bauer}},\ }\bibfield  {title} {\enquote {\bibinfo {title} {{Semiquantum}
  thermodynamics of complex ferrimagnets},}\ }\href {\doibase
  10.1103/physrevb.100.140401} {\bibfield  {journal} {\bibinfo  {journal}
  {Phys. Rev. B}\ }\textbf {\bibinfo {volume} {100}},\ \bibinfo {pages}
  {140401} (\bibinfo {year} {2019})}\BibitemShut {NoStop}%
\bibitem [{\citenamefont {Barker}\ \emph {et~al.}(2020)\citenamefont {Barker},
  \citenamefont {Pashov},\ and\ \citenamefont
  {Jackson}}]{Barker_ElectronStruct_2_044002_2020}%
  \BibitemOpen
  \bibfield  {author} {\bibinfo {author} {\bibfnamefont {Joseph}\ \bibnamefont
  {Barker}}, \bibinfo {author} {\bibfnamefont {Dimitar}\ \bibnamefont
  {Pashov}}, \ and\ \bibinfo {author} {\bibfnamefont {Jerome}\ \bibnamefont
  {Jackson}},\ }\bibfield  {title} {\enquote {\bibinfo {title} {{Electronic}
  structure and finite temperature magnetism of yttrium iron garnet},}\ }\href
  {\doibase 10.1088/2516-1075/abd097} {\bibfield  {journal} {\bibinfo
  {journal} {Electron. Struct.}\ }\textbf {\bibinfo {volume} {2}},\ \bibinfo
  {pages} {044002} (\bibinfo {year} {2020})}\BibitemShut {NoStop}%
\bibitem [{\citenamefont {Woo}\ \emph {et~al.}(2015)\citenamefont {Woo},
  \citenamefont {Wen}, \citenamefont {Semenov}, \citenamefont {Dudarev},\ and\
  \citenamefont {Ma}}]{Woo_PhysRevB_91_104306_2015}%
  \BibitemOpen
  \bibfield  {author} {\bibinfo {author} {\bibfnamefont {C.~H.}\ \bibnamefont
  {Woo}}, \bibinfo {author} {\bibfnamefont {Haohua}\ \bibnamefont {Wen}},
  \bibinfo {author} {\bibfnamefont {A.~A.}\ \bibnamefont {Semenov}}, \bibinfo
  {author} {\bibfnamefont {S.~L.}\ \bibnamefont {Dudarev}}, \ and\ \bibinfo
  {author} {\bibfnamefont {Pui-Wai}\ \bibnamefont {Ma}},\ }\bibfield  {title}
  {\enquote {\bibinfo {title} {{Quantum} heat bath for spin-lattice
  dynamics},}\ }\href {\doibase 10.1103/physrevb.91.104306} {\bibfield
  {journal} {\bibinfo  {journal} {Phys. Rev. B}\ }\textbf {\bibinfo {volume}
  {91}},\ \bibinfo {pages} {104306} (\bibinfo {year} {2015})}\BibitemShut
  {NoStop}%
\bibitem [{\citenamefont {Bergqvist}\ and\ \citenamefont
  {Bergman}(2018)}]{Bergqvist_PhysRevMater_2_013802_2018}%
  \BibitemOpen
  \bibfield  {author} {\bibinfo {author} {\bibfnamefont {Lars}\ \bibnamefont
  {Bergqvist}}\ and\ \bibinfo {author} {\bibfnamefont {Anders}\ \bibnamefont
  {Bergman}},\ }\bibfield  {title} {\enquote {\bibinfo {title} {{Realistic}
  finite temperature simulations of magnetic systems using quantum
  statistics},}\ }\href {\doibase 10.1103/physrevmaterials.2.013802} {\bibfield
   {journal} {\bibinfo  {journal} {Phys. Rev. Mater.}\ }\textbf {\bibinfo
  {volume} {2}},\ \bibinfo {pages} {013802} (\bibinfo {year}
  {2018})}\BibitemShut {NoStop}%
\bibitem [{\citenamefont {Evans}\ \emph {et~al.}(2015)\citenamefont {Evans},
  \citenamefont {Atxitia},\ and\ \citenamefont
  {Chantrell}}]{Evans_PhysRevB_91_144425_2015}%
  \BibitemOpen
  \bibfield  {author} {\bibinfo {author} {\bibfnamefont {R.~F.~L.}\
  \bibnamefont {Evans}}, \bibinfo {author} {\bibfnamefont {U.}~\bibnamefont
  {Atxitia}}, \ and\ \bibinfo {author} {\bibfnamefont {R.~W.}\ \bibnamefont
  {Chantrell}},\ }\bibfield  {title} {\enquote {\bibinfo {title}
  {{Quantitative} simulation of temperature-dependent magnetization dynamics
  and equilibrium properties of elemental ferromagnets},}\ }\href {\doibase
  10.1103/physrevb.91.144425} {\bibfield  {journal} {\bibinfo  {journal} {Phys.
  Rev. B}\ }\textbf {\bibinfo {volume} {91}},\ \bibinfo {pages} {144425}
  (\bibinfo {year} {2015})}\BibitemShut {NoStop}%
\bibitem [{\citenamefont {Anders}\ \emph {et~al.}(2022)\citenamefont {Anders},
  \citenamefont {Sait},\ and\ \citenamefont
  {Horsley}}]{Anders_NewJPhys_24_033020_2022}%
  \BibitemOpen
  \bibfield  {author} {\bibinfo {author} {\bibfnamefont {J}~\bibnamefont
  {Anders}}, \bibinfo {author} {\bibfnamefont {C~R~J}\ \bibnamefont {Sait}}, \
  and\ \bibinfo {author} {\bibfnamefont {S~A~R}\ \bibnamefont {Horsley}},\
  }\bibfield  {title} {\enquote {\bibinfo {title} {{Quantum} {Brownian} motion
  for magnets},}\ }\href {\doibase 10.1088/1367-2630/ac4ef2} {\bibfield
  {journal} {\bibinfo  {journal} {New J. Phys.}\ }\textbf {\bibinfo {volume}
  {24}},\ \bibinfo {pages} {033020} (\bibinfo {year} {2022})}\BibitemShut
  {NoStop}%
\bibitem [{\citenamefont {Walsh}\ \emph {et~al.}(2022)\citenamefont {Walsh},
  \citenamefont {Asta},\ and\ \citenamefont
  {Wang}}]{Walsh_npjComputMater_8_186_2022}%
  \BibitemOpen
  \bibfield  {author} {\bibinfo {author} {\bibfnamefont {Flynn}\ \bibnamefont
  {Walsh}}, \bibinfo {author} {\bibfnamefont {Mark}\ \bibnamefont {Asta}}, \
  and\ \bibinfo {author} {\bibfnamefont {Lin-Wang}\ \bibnamefont {Wang}},\
  }\bibfield  {title} {\enquote {\bibinfo {title} {{Realistic} magnetic
  thermodynamics by local quantization of a semiclassical {Heisenberg}
  model},}\ }\href {\doibase 10.1038/s41524-022-00875-8} {\bibfield  {journal}
  {\bibinfo  {journal} {npj Comput. Mater.}\ }\textbf {\bibinfo {volume} {8}},\
  \bibinfo {pages} {186} (\bibinfo {year} {2022})}\BibitemShut {NoStop}%
\bibitem [{\citenamefont {Shpyrko}\ \emph {et~al.}(2007)\citenamefont
  {Shpyrko}, \citenamefont {Isaacs}, \citenamefont {Logan}, \citenamefont
  {Feng}, \citenamefont {Aeppli}, \citenamefont {Jaramillo}, \citenamefont
  {Kim}, \citenamefont {Rosenbaum}, \citenamefont {Zschack}, \citenamefont
  {Sprung}, \citenamefont {Narayanan},\ and\ \citenamefont
  {Sandy}}]{Shpyrko_Nature_447_68_2007}%
  \BibitemOpen
  \bibfield  {author} {\bibinfo {author} {\bibfnamefont {O.~G.}\ \bibnamefont
  {Shpyrko}}, \bibinfo {author} {\bibfnamefont {E.~D.}\ \bibnamefont {Isaacs}},
  \bibinfo {author} {\bibfnamefont {J.~M.}\ \bibnamefont {Logan}}, \bibinfo
  {author} {\bibfnamefont {Yejun}\ \bibnamefont {Feng}}, \bibinfo {author}
  {\bibfnamefont {G.}~\bibnamefont {Aeppli}}, \bibinfo {author} {\bibfnamefont
  {R.}~\bibnamefont {Jaramillo}}, \bibinfo {author} {\bibfnamefont {H.~C.}\
  \bibnamefont {Kim}}, \bibinfo {author} {\bibfnamefont {T.~F.}\ \bibnamefont
  {Rosenbaum}}, \bibinfo {author} {\bibfnamefont {P.}~\bibnamefont {Zschack}},
  \bibinfo {author} {\bibfnamefont {M.}~\bibnamefont {Sprung}}, \bibinfo
  {author} {\bibfnamefont {S.}~\bibnamefont {Narayanan}}, \ and\ \bibinfo
  {author} {\bibfnamefont {A.~R.}\ \bibnamefont {Sandy}},\ }\bibfield  {title}
  {\enquote {\bibinfo {title} {{Direct} measurement of antiferromagnetic domain
  fluctuations},}\ }\href {\doibase 10.1038/nature05776} {\bibfield  {journal}
  {\bibinfo  {journal} {Nature}\ }\textbf {\bibinfo {volume} {447}},\ \bibinfo
  {pages} {68--71} (\bibinfo {year} {2007})}\BibitemShut {NoStop}%
\bibitem [{\citenamefont {Parrinello}\ and\ \citenamefont
  {Rahman}(1984)}]{Parrinello_JChemPhys_80_860_1984}%
  \BibitemOpen
  \bibfield  {author} {\bibinfo {author} {\bibfnamefont {M.}~\bibnamefont
  {Parrinello}}\ and\ \bibinfo {author} {\bibfnamefont {A.}~\bibnamefont
  {Rahman}},\ }\bibfield  {title} {\enquote {\bibinfo {title} {{Study} of an
  {F} center in molten {KCl}},}\ }\href {\doibase 10.1063/1.446740} {\bibfield
  {journal} {\bibinfo  {journal} {J. Chem. Phys.}\ }\textbf {\bibinfo {volume}
  {80}},\ \bibinfo {pages} {860--867} (\bibinfo {year} {1984})}\BibitemShut
  {NoStop}%
\bibitem [{\citenamefont {Habershon}\ \emph {et~al.}(2013)\citenamefont
  {Habershon}, \citenamefont {Manolopoulos}, \citenamefont {Markland},\ and\
  \citenamefont {Miller}}]{Habershon_AnnuRevPhysChem_64_387_2013}%
  \BibitemOpen
  \bibfield  {author} {\bibinfo {author} {\bibfnamefont {Scott}\ \bibnamefont
  {Habershon}}, \bibinfo {author} {\bibfnamefont {David~E.}\ \bibnamefont
  {Manolopoulos}}, \bibinfo {author} {\bibfnamefont {Thomas~E.}\ \bibnamefont
  {Markland}}, \ and\ \bibinfo {author} {\bibfnamefont {Thomas~F.}\
  \bibnamefont {Miller}},\ }\bibfield  {title} {\enquote {\bibinfo {title}
  {{Ring}-{Polymer} {Molecular} {Dynamics}: {Quantum} {Effects} in {Chemical}
  {Dynamics} from {Classical} {Trajectories} in an {Extended} {Phase}
  {Space}},}\ }\href {\doibase 10.1146/annurev-physchem-040412-110122}
  {\bibfield  {journal} {\bibinfo  {journal} {Annu. Rev. Phys. Chem.}\ }\textbf
  {\bibinfo {volume} {64}},\ \bibinfo {pages} {387--413} (\bibinfo {year}
  {2013})}\BibitemShut {NoStop}%
\bibitem [{\citenamefont {Runeson}\ and\ \citenamefont
  {Richardson}(2020)}]{Runeson_JChemPhys_152_084110_2020}%
  \BibitemOpen
  \bibfield  {author} {\bibinfo {author} {\bibfnamefont {Johan~E.}\
  \bibnamefont {Runeson}}\ and\ \bibinfo {author} {\bibfnamefont {Jeremy~O.}\
  \bibnamefont {Richardson}},\ }\bibfield  {title} {\enquote {\bibinfo {title}
  {{Generalized} spin mapping for quantum-classical dynamics},}\ }\href
  {\doibase 10.1063/1.5143412} {\bibfield  {journal} {\bibinfo  {journal} {J.
  Chem. Phys.}\ }\textbf {\bibinfo {volume} {152}},\ \bibinfo {pages} {084110}
  (\bibinfo {year} {2020})}\BibitemShut {NoStop}%
\bibitem [{\citenamefont {Coronado}(2019)}]{Coronado_NatRevMater_5_87_2019}%
  \BibitemOpen
  \bibfield  {author} {\bibinfo {author} {\bibfnamefont {Eugenio}\ \bibnamefont
  {Coronado}},\ }\bibfield  {title} {\enquote {\bibinfo {title} {{Molecular}
  magnetism: from chemical design to spin control in molecules, materials and
  devices},}\ }\href {\doibase 10.1038/s41578-019-0146-8} {\bibfield  {journal}
  {\bibinfo  {journal} {Nat. Rev. Mater.}\ }\textbf {\bibinfo {volume} {5}},\
  \bibinfo {pages} {87--104} (\bibinfo {year} {2019})}\BibitemShut {NoStop}%
\bibitem [{\citenamefont {Bossion}\ \emph {et~al.}(2022)\citenamefont
  {Bossion}, \citenamefont {Ying}, \citenamefont {Chowdhury},\ and\
  \citenamefont {Huo}}]{Bossion_JChemPhys_157_084105_2022}%
  \BibitemOpen
  \bibfield  {author} {\bibinfo {author} {\bibfnamefont {Duncan}\ \bibnamefont
  {Bossion}}, \bibinfo {author} {\bibfnamefont {Wenxiang}\ \bibnamefont
  {Ying}}, \bibinfo {author} {\bibfnamefont {Sutirtha~N.}\ \bibnamefont
  {Chowdhury}}, \ and\ \bibinfo {author} {\bibfnamefont {Pengfei}\ \bibnamefont
  {Huo}},\ }\bibfield  {title} {\enquote {\bibinfo {title} {{Non}-adiabatic
  mapping dynamics in the phase space of the {$SU(N)$} {Lie} group},}\ }\href
  {\doibase 10.1063/5.0094893} {\bibfield  {journal} {\bibinfo  {journal} {J.
  Chem. Phys.}\ }\textbf {\bibinfo {volume} {157}},\ \bibinfo {pages} {084105}
  (\bibinfo {year} {2022})}\BibitemShut {NoStop}%
\bibitem [{\citenamefont {Zhang}\ and\ \citenamefont
  {Batista}(2021)}]{Zhang_PhysRevB_104_104409_2021}%
  \BibitemOpen
  \bibfield  {author} {\bibinfo {author} {\bibfnamefont {Hao}\ \bibnamefont
  {Zhang}}\ and\ \bibinfo {author} {\bibfnamefont {Cristian~D.}\ \bibnamefont
  {Batista}},\ }\bibfield  {title} {\enquote {\bibinfo {title} {{Classical}
  spin dynamics based on {SU}({N}) coherent states},}\ }\href {\doibase
  10.1103/physrevb.104.104409} {\bibfield  {journal} {\bibinfo  {journal}
  {Phys. Rev. B}\ }\textbf {\bibinfo {volume} {104}},\ \bibinfo {pages}
  {104409} (\bibinfo {year} {2021})}\BibitemShut {NoStop}%
\bibitem [{\citenamefont
  {Kochetov}(1998)}]{kochetovQuasiclassicalPathIntegral1998}%
  \BibitemOpen
  \bibfield  {author} {\bibinfo {author} {\bibfnamefont {E.~A.}\ \bibnamefont
  {Kochetov}},\ }\bibfield  {title} {\enquote {\bibinfo {title} {Quasiclassical
  path integral in coherent-state manifolds},}\ }\href {\doibase
  10.1088/0305-4470/31/19/013} {\bibfield  {journal} {\bibinfo  {journal}
  {Journal of Physics A: Mathematical and General}\ }\textbf {\bibinfo {volume}
  {31}},\ \bibinfo {pages} {4473} (\bibinfo {year} {1998})}\BibitemShut
  {NoStop}%
\bibitem [{\citenamefont {Cabra}\ \emph {et~al.}(1997)\citenamefont {Cabra},
  \citenamefont {Dobry}, \citenamefont {Greco},\ and\ \citenamefont
  {Rossini}}]{cabraPathIntegralRepresentation1997}%
  \BibitemOpen
  \bibfield  {author} {\bibinfo {author} {\bibfnamefont {Daniel~C.}\
  \bibnamefont {Cabra}}, \bibinfo {author} {\bibfnamefont {Ariel}\ \bibnamefont
  {Dobry}}, \bibinfo {author} {\bibfnamefont {Andr{\'e}s}\ \bibnamefont
  {Greco}}, \ and\ \bibinfo {author} {\bibfnamefont {Gerardo~L.}\ \bibnamefont
  {Rossini}},\ }\bibfield  {title} {\enquote {\bibinfo {title} {On the path
  integral representation for spin systems},}\ }\href {\doibase
  10.1088/0305-4470/30/8/016} {\bibfield  {journal} {\bibinfo  {journal}
  {Journal of Physics A: Mathematical and General}\ }\textbf {\bibinfo {volume}
  {30}},\ \bibinfo {pages} {2699--2704} (\bibinfo {year} {1997})}\BibitemShut
  {NoStop}%
\bibitem [{\citenamefont
  {Klauder}(1979)}]{klauderPathIntegralsStationaryphase1979}%
  \BibitemOpen
  \bibfield  {author} {\bibinfo {author} {\bibfnamefont {John~R.}\ \bibnamefont
  {Klauder}},\ }\bibfield  {title} {\enquote {\bibinfo {title} {Path integrals
  and stationary-phase approximations},}\ }\href {\doibase
  10.1103/PhysRevD.19.2349} {\bibfield  {journal} {\bibinfo  {journal}
  {Physical Review D}\ }\textbf {\bibinfo {volume} {19}},\ \bibinfo {pages}
  {2349--2356} (\bibinfo {year} {1979})}\BibitemShut {NoStop}%
\bibitem [{\citenamefont {Gerlach}\ and\ \citenamefont
  {Stern}(1922)}]{gerlachExperimentelleNachweisRichtungsquantelung1922}%
  \BibitemOpen
  \bibfield  {author} {\bibinfo {author} {\bibfnamefont {Walther}\ \bibnamefont
  {Gerlach}}\ and\ \bibinfo {author} {\bibfnamefont {Otto}\ \bibnamefont
  {Stern}},\ }\bibfield  {title} {\enquote {\bibinfo {title} {{Der
  experimentelle Nachweis der Richtungsquantelung im Magnetfeld}},}\ }\href
  {\doibase 10.1007/BF01326983} {\bibfield  {journal} {\bibinfo  {journal}
  {Zeitschrift f\"ur Physik}\ }\textbf {\bibinfo {volume} {9}},\ \bibinfo
  {pages} {349--352} (\bibinfo {year} {1922})}\BibitemShut {NoStop}%
\bibitem [{\citenamefont
  {Radcliffe}(1971)}]{Radcliffe_JPhysAGenPhys_4_313_1971}%
  \BibitemOpen
  \bibfield  {author} {\bibinfo {author} {\bibfnamefont {J~M}\ \bibnamefont
  {Radcliffe}},\ }\bibfield  {title} {\enquote {\bibinfo {title} {{Some}
  properties of coherent spin states},}\ }\href {\doibase
  10.1088/0305-4470/4/3/009} {\bibfield  {journal} {\bibinfo  {journal} {J.
  Phys. A: Gen. Phys.}\ }\textbf {\bibinfo {volume} {4}},\ \bibinfo {pages}
  {313--323} (\bibinfo {year} {1971})}\BibitemShut {NoStop}%
\bibitem [{\citenamefont {Lee~Loh}\ and\ \citenamefont
  {Kim}(2015)}]{Lee_Loh_AmJPhys_83_30_2015}%
  \BibitemOpen
  \bibfield  {author} {\bibinfo {author} {\bibfnamefont {Yen}\ \bibnamefont
  {Lee~Loh}}\ and\ \bibinfo {author} {\bibfnamefont {Monica}\ \bibnamefont
  {Kim}},\ }\bibfield  {title} {\enquote {\bibinfo {title} {{Visualizing} spin
  states using the spin coherent state representation},}\ }\href {\doibase
  10.1119/1.4898595} {\bibfield  {journal} {\bibinfo  {journal} {Am. J. Phys.}\
  }\textbf {\bibinfo {volume} {83}},\ \bibinfo {pages} {30--35} (\bibinfo
  {year} {2015})}\BibitemShut {NoStop}%
\bibitem [{\citenamefont {Stone}(1989)}]{Stone_NuclPhysB_314_557_1989}%
  \BibitemOpen
  \bibfield  {author} {\bibinfo {author} {\bibfnamefont {Michael}\ \bibnamefont
  {Stone}},\ }\bibfield  {title} {\enquote {\bibinfo {title} {{Supersymmetry}
  and the quantum mechanics of spin},}\ }\href {\doibase
  10.1016/0550-3213(89)90408-2} {\bibfield  {journal} {\bibinfo  {journal}
  {Nucl. Phys. B}\ }\textbf {\bibinfo {volume} {314}},\ \bibinfo {pages}
  {557--586} (\bibinfo {year} {1989})}\BibitemShut {NoStop}%
\bibitem [{\citenamefont {Stone}\ \emph {et~al.}(2000)\citenamefont {Stone},
  \citenamefont {Park},\ and\ \citenamefont
  {Garg}}]{Stone_JMathPhys_41_8025_2000}%
  \BibitemOpen
  \bibfield  {author} {\bibinfo {author} {\bibfnamefont {Michael}\ \bibnamefont
  {Stone}}, \bibinfo {author} {\bibfnamefont {Kee-Su}\ \bibnamefont {Park}}, \
  and\ \bibinfo {author} {\bibfnamefont {Anupam}\ \bibnamefont {Garg}},\
  }\bibfield  {title} {\enquote {\bibinfo {title} {{The} semiclassical
  propagator for spin coherent states},}\ }\href {\doibase 10.1063/1.1320856}
  {\bibfield  {journal} {\bibinfo  {journal} {J. Math. Phys.}\ }\textbf
  {\bibinfo {volume} {41}},\ \bibinfo {pages} {8025--8049} (\bibinfo {year}
  {2000})}\BibitemShut {NoStop}%
\bibitem [{\citenamefont {Koh}(2018)}]{Koh_PhysRevB_97_094417_2018}%
  \BibitemOpen
  \bibfield  {author} {\bibinfo {author} {\bibfnamefont {Yang~Wei}\
  \bibnamefont {Koh}},\ }\bibfield  {title} {\enquote {\bibinfo {title}
  {{Effects} of dynamical paths on the energy gap and the corrections to the
  free energy in path integrals of mean-field quantum spin systems},}\ }\href
  {\doibase 10.1103/physrevb.97.094417} {\bibfield  {journal} {\bibinfo
  {journal} {Phys. Rev. B}\ }\textbf {\bibinfo {volume} {97}},\ \bibinfo
  {pages} {094417} (\bibinfo {year} {2018})}\BibitemShut {NoStop}%
\bibitem [{Note1()}]{Note1}%
  \BibitemOpen
  \bibinfo {note} {Of course one can also define these in terms of the raising
  operator or any linear combination of these \cite
  {Nemoto_JPhysAMathGen_33_3493_2000}}\BibitemShut {NoStop}%
\bibitem [{\citenamefont
  {Karchev}(2012)}]{karchevPathIntegralRepresentation2012}%
  \BibitemOpen
  \bibfield  {author} {\bibinfo {author} {\bibfnamefont {Naoum}\ \bibnamefont
  {Karchev}},\ }\bibfield  {title} {\enquote {\bibinfo {title} {Path integral
  representation for spin systens},}\ }\href@noop {} {\bibfield  {journal}
  {\bibinfo  {journal} {arXiv:1211.4509 [cond-mat]}\ } (\bibinfo {year}
  {2012})},\ \Eprint {http://arxiv.org/abs/1211.4509} {arXiv:1211.4509
  [cond-mat]} \BibitemShut {NoStop}%
\bibitem [{\citenamefont {Lieb}(1973)}]{liebClassicalLimitQuantum1973}%
  \BibitemOpen
  \bibfield  {author} {\bibinfo {author} {\bibfnamefont {Elliott~H.}\
  \bibnamefont {Lieb}},\ }\bibfield  {title} {\enquote {\bibinfo {title} {The
  classical limit of quantum spin systems},}\ }\href {\doibase
  10.1007/BF01646493} {\bibfield  {journal} {\bibinfo  {journal}
  {Communications in Mathematical Physics}\ }\textbf {\bibinfo {volume} {31}},\
  \bibinfo {pages} {327--340} (\bibinfo {year} {1973})}\BibitemShut {NoStop}%
\bibitem [{\citenamefont {Conlon}\ and\ \citenamefont
  {Solovej}(1990)}]{conlonAsymptoticLimitsQuantum1990}%
  \BibitemOpen
  \bibfield  {author} {\bibinfo {author} {\bibfnamefont {J.~G.}\ \bibnamefont
  {Conlon}}\ and\ \bibinfo {author} {\bibfnamefont {J.~P.}\ \bibnamefont
  {Solovej}},\ }\bibfield  {title} {\enquote {\bibinfo {title} {On asymptotic
  limits for the quantum {{Heisenberg}} model},}\ }\href {\doibase
  10.1088/0305-4470/23/14/018} {\bibfield  {journal} {\bibinfo  {journal}
  {Journal of Physics A: Mathematical and General}\ }\textbf {\bibinfo {volume}
  {23}},\ \bibinfo {pages} {3199} (\bibinfo {year} {1990})}\BibitemShut
  {NoStop}%
\bibitem [{\citenamefont {Millard}\ and\ \citenamefont
  {Leff}(2003)}]{millardInfiniteSpinLimit2003}%
  \BibitemOpen
  \bibfield  {author} {\bibinfo {author} {\bibfnamefont {Kenneth}\ \bibnamefont
  {Millard}}\ and\ \bibinfo {author} {\bibfnamefont {Harvey~S.}\ \bibnamefont
  {Leff}},\ }\bibfield  {title} {\enquote {\bibinfo {title} {Infinite-{{Spin
  Limit}} of the {{Quantum Heisenberg Model}}},}\ }\href {\doibase
  10.1063/1.1665664} {\bibfield  {journal} {\bibinfo  {journal} {Journal of
  Mathematical Physics}\ }\textbf {\bibinfo {volume} {12}},\ \bibinfo {pages}
  {1000--1005} (\bibinfo {year} {2003})}\BibitemShut {NoStop}%
\bibitem [{\citenamefont {'t~Hooft}(1974)}]{hooftPlanarDiagramTheory1974}%
  \BibitemOpen
  \bibfield  {author} {\bibinfo {author} {\bibfnamefont {G.}~\bibnamefont
  {'t~Hooft}},\ }\bibfield  {title} {\enquote {\bibinfo {title} {A planar
  diagram theory for strong interactions},}\ }\href {\doibase
  10.1016/0550-3213(74)90154-0} {\bibfield  {journal} {\bibinfo  {journal}
  {Nuclear Physics B}\ }\textbf {\bibinfo {volume} {72}},\ \bibinfo {pages}
  {461--473} (\bibinfo {year} {1974})}\BibitemShut {NoStop}%
\bibitem [{\citenamefont {Greiner}\ \emph {et~al.}(2000)\citenamefont
  {Greiner}, \citenamefont {Neise},\ and\ \citenamefont
  {St{\"o}cker}}]{greinerThermodynamicsStatisticalMechanics2000}%
  \BibitemOpen
  \bibfield  {author} {\bibinfo {author} {\bibfnamefont {Walter}\ \bibnamefont
  {Greiner}}, \bibinfo {author} {\bibfnamefont {Ludwig}\ \bibnamefont {Neise}},
  \ and\ \bibinfo {author} {\bibfnamefont {Horst}\ \bibnamefont
  {St{\"o}cker}},\ }\href@noop {} {\emph {\bibinfo {title} {Thermodynamics and
  {{Statistical Mechanics}}}}}\ (\bibinfo  {publisher} {{Springer New York}},\
  \bibinfo {year} {2000})\BibitemShut {NoStop}%
\bibitem [{\citenamefont {Deymier}\ \emph {et~al.}(2016)\citenamefont
  {Deymier}, \citenamefont {Runge},\ and\ \citenamefont
  {Muralidharan}}]{deymierMultiscaleParadigmsIntegrated2016}%
  \BibitemOpen
  \bibinfo {editor} {\bibfnamefont {Pierre}\ \bibnamefont {Deymier}}, \bibinfo
  {editor} {\bibfnamefont {Keith}\ \bibnamefont {Runge}}, \ and\ \bibinfo
  {editor} {\bibfnamefont {Krishna}\ \bibnamefont {Muralidharan}},\ eds.,\
  \href {\doibase 10.1007/978-3-319-24529-4} {\emph {\bibinfo {title}
  {Multiscale {{Paradigms}} in {{Integrated Computational Materials Science}}
  and {{Engineering}}}}},\ \bibinfo {series} {Springer {{Series}} in
  {{Materials Science}}}, Vol.\ \bibinfo {volume} {226}\ (\bibinfo  {publisher}
  {{Springer International Publishing}},\ \bibinfo {address} {{Cham}},\
  \bibinfo {year} {2016})\BibitemShut {NoStop}%
\bibitem [{\citenamefont {Halilov}\ \emph {et~al.}(1998)\citenamefont
  {Halilov}, \citenamefont {Eschrig}, \citenamefont {Perlov},\ and\
  \citenamefont {Oppeneer}}]{Halilov_PhysRevB_58_293_1998}%
  \BibitemOpen
  \bibfield  {author} {\bibinfo {author} {\bibfnamefont {S.~V.}\ \bibnamefont
  {Halilov}}, \bibinfo {author} {\bibfnamefont {H.}~\bibnamefont {Eschrig}},
  \bibinfo {author} {\bibfnamefont {A.~Y.}\ \bibnamefont {Perlov}}, \ and\
  \bibinfo {author} {\bibfnamefont {P.~M.}\ \bibnamefont {Oppeneer}},\
  }\bibfield  {title} {\enquote {\bibinfo {title} {{Adiabatic} spin dynamics
  from spin-density-functional theory: {Application} to {Fe}, {Co}, and
  {Ni}},}\ }\href {\doibase 10.1103/physrevb.58.293} {\bibfield  {journal}
  {\bibinfo  {journal} {Phys. Rev. B}\ }\textbf {\bibinfo {volume} {58}},\
  \bibinfo {pages} {293--302} (\bibinfo {year} {1998})}\BibitemShut {NoStop}%
\bibitem [{\citenamefont {Chubykalo}\ \emph {et~al.}(2003)\citenamefont
  {Chubykalo}, \citenamefont {Smirnov-Rueda}, \citenamefont {Gonzalez},
  \citenamefont {Wongsam}, \citenamefont {Chantrell},\ and\ \citenamefont
  {Nowak}}]{Chubykalo_JMagnMagnMater_266_28_2003}%
  \BibitemOpen
  \bibfield  {author} {\bibinfo {author} {\bibfnamefont {O.}~\bibnamefont
  {Chubykalo}}, \bibinfo {author} {\bibfnamefont {R.}~\bibnamefont
  {Smirnov-Rueda}}, \bibinfo {author} {\bibfnamefont {J.M.}\ \bibnamefont
  {Gonzalez}}, \bibinfo {author} {\bibfnamefont {M.A.}\ \bibnamefont
  {Wongsam}}, \bibinfo {author} {\bibfnamefont {R.W.}\ \bibnamefont
  {Chantrell}}, \ and\ \bibinfo {author} {\bibfnamefont {U.}~\bibnamefont
  {Nowak}},\ }\bibfield  {title} {\enquote {\bibinfo {title} {{Brownian}
  dynamics approach to interacting magnetic moments},}\ }\href {\doibase
  10.1016/s0304-8853(03)00452-9} {\bibfield  {journal} {\bibinfo  {journal} {J.
  Magn. Magn. Mater.}\ }\textbf {\bibinfo {volume} {266}},\ \bibinfo {pages}
  {28--35} (\bibinfo {year} {2003})}\BibitemShut {NoStop}%
\bibitem [{\citenamefont {Mryasov}\ \emph {et~al.}(2005)\citenamefont
  {Mryasov}, \citenamefont {Nowak}, \citenamefont {Guslienko},\ and\
  \citenamefont {Chantrell}}]{Mryasov_EurLett_EPL_69_805_2005}%
  \BibitemOpen
  \bibfield  {author} {\bibinfo {author} {\bibfnamefont {O.~N}\ \bibnamefont
  {Mryasov}}, \bibinfo {author} {\bibfnamefont {U}~\bibnamefont {Nowak}},
  \bibinfo {author} {\bibfnamefont {K.~Y}\ \bibnamefont {Guslienko}}, \ and\
  \bibinfo {author} {\bibfnamefont {R.~W}\ \bibnamefont {Chantrell}},\
  }\bibfield  {title} {\enquote {\bibinfo {title} {{Temperature}-dependent
  magnetic properties of {FePt}: {Effective} spin {Hamiltonian} model},}\
  }\href {\doibase 10.1209/epl/i2004-10404-2} {\bibfield  {journal} {\bibinfo
  {journal} {Eur. Lett. (EPL)}\ }\textbf {\bibinfo {volume} {69}},\ \bibinfo
  {pages} {805--811} (\bibinfo {year} {2005})}\BibitemShut {NoStop}%
\bibitem [{\citenamefont {Skubic}\ \emph {et~al.}(2008)\citenamefont {Skubic},
  \citenamefont {Hellsvik}, \citenamefont {Nordstr{\"o}m},\ and\ \citenamefont
  {Eriksson}}]{Skubic_JPhysCondensMatter_20_315203_2008}%
  \BibitemOpen
  \bibfield  {author} {\bibinfo {author} {\bibfnamefont {B}~\bibnamefont
  {Skubic}}, \bibinfo {author} {\bibfnamefont {J}~\bibnamefont {Hellsvik}},
  \bibinfo {author} {\bibfnamefont {L}~\bibnamefont {Nordstr{\"o}m}}, \ and\
  \bibinfo {author} {\bibfnamefont {O}~\bibnamefont {Eriksson}},\ }\bibfield
  {title} {\enquote {\bibinfo {title} {{A} method for atomistic spin dynamics
  simulations: implementation and examples},}\ }\href {\doibase
  10.1088/0953-8984/20/31/315203} {\bibfield  {journal} {\bibinfo  {journal}
  {J. Phys.: Condens. Matter}\ }\textbf {\bibinfo {volume} {20}},\ \bibinfo
  {pages} {315203} (\bibinfo {year} {2008})}\BibitemShut {NoStop}%
\bibitem [{\citenamefont {Evans}\ \emph {et~al.}(2014)\citenamefont {Evans},
  \citenamefont {Fan}, \citenamefont {Chureemart}, \citenamefont {Ostler},
  \citenamefont {Ellis},\ and\ \citenamefont
  {Chantrell}}]{Evans_JPhysCondensMatter_26_103202_2014}%
  \BibitemOpen
  \bibfield  {author} {\bibinfo {author} {\bibfnamefont {R~F~L}\ \bibnamefont
  {Evans}}, \bibinfo {author} {\bibfnamefont {W~J}\ \bibnamefont {Fan}},
  \bibinfo {author} {\bibfnamefont {P}~\bibnamefont {Chureemart}}, \bibinfo
  {author} {\bibfnamefont {T~A}\ \bibnamefont {Ostler}}, \bibinfo {author}
  {\bibfnamefont {M~O~A}\ \bibnamefont {Ellis}}, \ and\ \bibinfo {author}
  {\bibfnamefont {R~W}\ \bibnamefont {Chantrell}},\ }\bibfield  {title}
  {\enquote {\bibinfo {title} {{Atomistic} spin model simulations of magnetic
  nanomaterials},}\ }\href {\doibase 10.1088/0953-8984/26/10/103202} {\bibfield
   {journal} {\bibinfo  {journal} {J. Phys.: Condens. Matter}\ }\textbf
  {\bibinfo {volume} {26}},\ \bibinfo {pages} {103202} (\bibinfo {year}
  {2014})}\BibitemShut {NoStop}%
\bibitem [{\citenamefont {Chen}\ and\ \citenamefont
  {Kim}(2004)}]{chenBrownianDynamicsMolecular2004}%
  \BibitemOpen
  \bibfield  {author} {\bibinfo {author} {\bibfnamefont {Jim~C.}\ \bibnamefont
  {Chen}}\ and\ \bibinfo {author} {\bibfnamefont {Albert~S.}\ \bibnamefont
  {Kim}},\ }\bibfield  {title} {\enquote {\bibinfo {title} {Brownian
  {{Dynamics}}, {{Molecular Dynamics}}, and {{Monte Carlo}} modeling of
  colloidal systems},}\ }\href {\doibase 10.1016/j.cis.2004.10.001} {\bibfield
  {journal} {\bibinfo  {journal} {Advances in Colloid and Interface Science}\
  }\textbf {\bibinfo {volume} {112}},\ \bibinfo {pages} {159--173} (\bibinfo
  {year} {2004})}\BibitemShut {NoStop}%
\bibitem [{\citenamefont {Thibaudeau}\ and\ \citenamefont
  {Beaujouan}(2012)}]{Thibaudeau_PhysAStatMechitsAppl_391_1963_2012}%
  \BibitemOpen
  \bibfield  {author} {\bibinfo {author} {\bibfnamefont {Pascal}\ \bibnamefont
  {Thibaudeau}}\ and\ \bibinfo {author} {\bibfnamefont {David}\ \bibnamefont
  {Beaujouan}},\ }\bibfield  {title} {\enquote {\bibinfo {title}
  {{Thermostatting} the atomic spin dynamics from controlled demons},}\ }\href
  {\doibase 10.1016/j.physa.2011.11.030} {\bibfield  {journal} {\bibinfo
  {journal} {Phys. A: Stat. Mech. its Appl.}\ }\textbf {\bibinfo {volume}
  {391}},\ \bibinfo {pages} {1963--1971} (\bibinfo {year} {2012})}\BibitemShut
  {NoStop}%
\bibitem [{\citenamefont {Kuz'min}(2005)}]{Kuzmin_PhysRevLett_94_107204_2005}%
  \BibitemOpen
  \bibfield  {author} {\bibinfo {author} {\bibfnamefont {M.~D.}\ \bibnamefont
  {Kuz'min}},\ }\bibfield  {title} {\enquote {\bibinfo {title} {{Shape} of
  {Temperature} {Dependence} of {Spontaneous} {Magnetization} of
  {Ferromagnets}: {Quantitative} {Analysis}},}\ }\href {\doibase
  10.1103/physrevlett.94.107204} {\bibfield  {journal} {\bibinfo  {journal}
  {Phys. Rev. Lett.}\ }\textbf {\bibinfo {volume} {94}},\ \bibinfo {pages}
  {107204} (\bibinfo {year} {2005})}\BibitemShut {NoStop}%
\bibitem [{\citenamefont {Zwanziger}\ \emph {et~al.}(1990)\citenamefont
  {Zwanziger}, \citenamefont {Koenig},\ and\ \citenamefont
  {Pines}}]{zwanzigerBerryPhase1990}%
  \BibitemOpen
  \bibfield  {author} {\bibinfo {author} {\bibfnamefont {J~W}\ \bibnamefont
  {Zwanziger}}, \bibinfo {author} {\bibfnamefont {M}~\bibnamefont {Koenig}}, \
  and\ \bibinfo {author} {\bibfnamefont {A}~\bibnamefont {Pines}},\ }\bibfield
  {title} {\enquote {\bibinfo {title} {Berry's {{Phase}}},}\ }\href {\doibase
  10.1146/annurev.pc.41.100190.003125} {\bibfield  {journal} {\bibinfo
  {journal} {Annual Review of Physical Chemistry}\ }\textbf {\bibinfo {volume}
  {41}},\ \bibinfo {pages} {601--646} (\bibinfo {year} {1990})}\BibitemShut
  {NoStop}%
\bibitem [{\citenamefont {Nussle}\ \emph {et~al.}(2023)\citenamefont {Nussle},
  \citenamefont {Nicolis},\ and\ \citenamefont {Barker}}]{thisDataset}%
  \BibitemOpen
  \bibfield  {author} {\bibinfo {author} {\bibfnamefont {Thomas}\ \bibnamefont
  {Nussle}}, \bibinfo {author} {\bibfnamefont {Stam}\ \bibnamefont {Nicolis}},
  \ and\ \bibinfo {author} {\bibfnamefont {Joseph}\ \bibnamefont {Barker}},\
  }\href {\doibase 10.5281/zenodo.7692092} {\enquote {\bibinfo {title}
  {{Sources for: Numerical Simulations of a Spin Dynamics Model Based on a Path
  Integral Approach (v1.0.5) [Data set]}},}\ } (\bibinfo {year} {2023}),\
  \bibinfo {note} {{Zenodo}}\BibitemShut {NoStop}%
\bibitem [{\citenamefont {Nemoto}(2000)}]{Nemoto_JPhysAMathGen_33_3493_2000}%
  \BibitemOpen
  \bibfield  {author} {\bibinfo {author} {\bibfnamefont {Kae}\ \bibnamefont
  {Nemoto}},\ }\bibfield  {title} {\enquote {\bibinfo {title} {{Generalized}
  coherent states for {$SU(n)$} systems},}\ }\href {\doibase
  10.1088/0305-4470/33/17/307} {\bibfield  {journal} {\bibinfo  {journal} {J.
  Phys. A: Math. Gen.}\ }\textbf {\bibinfo {volume} {33}},\ \bibinfo {pages}
  {3493--3506} (\bibinfo {year} {2000})}\BibitemShut {NoStop}%
\end{thebibliography}%

\end{document}